\documentclass{aa520}
\usepackage{txfonts}
\usepackage{graphicx}
\usepackage{natbib}

\def\kms{{\rm{km}}/{\rm{s}}}

\def\msun{\rm{M_{\odot}}}
\def\msunpc3{\rm{M_{\odot} pc^{-3}}}

\def\kpc{\rm{kpc}}
\def\rhoc{\rho_{\rm{cld}}}

\def\rhopsf{\dot{\rho}_{\rm{sf}}}
\def\csfr{c_{\rm{sf}}}

\def\Qeff{Q_{\rm{eff}}}
\def\Qgas{Q_{\rm{gas}}}
\def\Qstr{Q_{\rm{star}}}

\def\ek{\epsilon}
\def\etac{\eta_{\rm{coll}}}

\def\gl#1#2{\begin{equation}\label{#1}#2\end{equation}}
\def\dd#1#2{\frac{\partial {#1}}{\partial {#2}}}

\begin{document}

\title{Gas Physics, Disk Fragmentation, and \\ Bulge Formation in Young Galaxies}

\author{Andreas Immeli\inst{1}, Markus Samland\inst{1}, Ortwin Gerhard\inst{1},
        \and Pieter Westera\inst{2}}

\institute{Astronomisches Institut der Universit{\"a}t Basel,
           Venusstrasse 7, CH-4102 Binningen
           \and
           Observat{\'o}rio do Valongo, Universidade Federal do Rio de Janeiro,
           Ladeira do Pedro Ant{\^o}nio, 43, CEP 20080-090, Rio de Janeiro, Brazil}

\date{Received / Accepted}

\authorrunning{Immeli et al.}

%
\abstract{
We investigate the evolution of star-forming gas-rich disks, using a
3D chemodynamical model including a dark halo, stars, and a two-phase
interstellar medium with feedback processes from the stars. We show
that galaxy evolution proceeds along very different routes depending
on whether it is the gas disk or the stellar disk which first becomes
unstable, as measured by the respective Q-parameters. This in turn
depends on the uncertain efficiency of energy dissipation of the cold
cloud component from which stars form.

When the cold gas cools efficiently and drives the instability, the
galactic disk fragments and forms a number of massive clumps of stars
and gas. The clumps spiral to the center of the galaxy in a few
dynamical times and merge there to form a central bulge component in a
strong starburst.  When the kinetic energy of the cold clouds is
dissipated at a lower rate, stars form from the gas in a more
quiescent mode, and an instability only sets in at later times, when
the surface density of the stellar disk has grown sufficiently high.
The system then forms a stellar bar, which channels gas into the
center, evolves, and forms a bulge whose stars are the result of a
more extended star formation history.

We investigate the stability of the gas-stellar disks in both regimes,
as well as the star formation rates and element enrichment.  We study
the morphology of the evolving disks, calculating spatially resolved
colours from the distribution of stars in age and metallicity,
including dust absorption. We then discuss morphological observations
such as clumpy structures and chain galaxies at high redshift as
possible signatures of fragmenting, gas-rich disks. Finally, we
investigate abundance ratio distributions as a means to distinguish
the different scenarios for bulge formation.
\keywords{
ISM: kinematics and dynamics -- 
ISM: structure --
Galaxies: abundances
-- Galaxies: bulges
-- Galaxies: evolution
-- Galaxies: kinematics and dynamics
}
}

\maketitle

%
\section{Introduction}


The formation of galaxies is one of the important questions in current
astrophysical research.  With HST and earth-bound 10m class telescopes
it is now possible to study the evolution of galaxies with redshift by
direct observation. Although no complete picture is available at the
moment, some facts are well established.
Surveys undertaken around $z \simeq 1$ have shown that luminous
elliptical and spiral galaxies were largely formed by then, and have
only moderately evolved to the present time \citep{brinchmann98,
  lilly98, abraham99, abraham99b, abraham00, dickinson00}.  Beyond
$z=1.4$, there are far fewer high-luminosity galaxies off all types
compared to low redshifts (\citealt{dickinson00}, see also
\citealt{driver98}). Thus the bulge-to-disk Hubble sequence of
galaxies appears to have formed at redshifts slightly beyond $z \sim
1$ \citep{abraham00, kajisawa01}, although part of the stellar
populations of these galaxies might have formed earlier.

At $z \simeq 0.5$ spiral galaxies and barred galaxies are observed,
and it seems that by then the full Hubble sequence is in place, quite
similar to the galaxy distribution today. However, the frequency of
barred galaxies drops sharply beyond $z \simeq 0.5$
\citep[e.g.,][]{abraham99}.  \citet{vdb02} showed that this is not an
selection effect. They shifted a local galaxy sample to higher
redshifts and concluded that most of the barred galaxies would still
be visible. Thus, the absence of bars at higher redshifts seems to be
real.

Already at $z\sim 1$ around $30\%$ of galaxies are morphologically
peculiar. At higher redshifts many clumpy structures and compact
objects are observed. Most of these cannot be attached to the
traditional Hubble scheme, but show irregular morphologies
\citep[e.g.,][]{abraham96, vandenbergh00}. The anomalous morphologies
observed in these high redshift objects cannot be explained through
band-shifting effects alone, because the irregularities persist also
in NICMOS observations \citep{dickinson00}, probing the visual
restframe wavelength of these objects.


Observations of high redshift galaxies currently provide information
only about the global properties of these objects. Detailed data like
stellar metallicity distributions, stellar kinematics, or gas distributions
are only available for local galaxies.  
To understand galaxy formation in a consistent picture, models must be
developed that can be compared with observations over the whole
observed redshift range. These models should be able to explain the
properties of distant galaxies as well as the detailed data on local
galaxies.


With high resolution cosmological simulations, large progress has been
made in understanding cosmic structure formation
\citep[e.g.,][]{navarro96, moore98, jenkins01, klypin01}. However, on
galactic scales these simulations still lack the necessary resolution
to describe the processes relevant for baryon dissipation and star
formation.  Therefore two approaches for describing galaxy formation
and evolution have been developed.  Semi-analytical modeling, based on
simple assumptions to describe the baryonic physics and star formation
in the dark halos that form in the cosmological simulations, has been
used to analyze the global properties of galaxy samples 
\citep[e.g.,][]{kauffmann93, guiderdoni98, cole00}.

On the other hand, dynamical models, using a subset of the
cosmological information as initial conditions on smaller scales, have
been used to investigate the detailed structure of forming galaxies
\citep[e.g.,][]{steinmetz95, navarro97b, sommer99, williams01, samland03}.  
These small-scale dynamical models still need to
describe star formation with a simple parametrization, but they
contain a much more detailed description of the dynamics, feedback,
and, in some cases, element enrichment. Thus with these models it is
possible to predict observable properties through the galaxy assembly
process, and to compare directly with observations of high redshift
galaxies \citep[e.g.,][]{contardo98, westera02, abadi03}. 
Because these models can be
calculated over a Hubble time, one can also directly compare detailed
present-day characteristics (e.g., metallicities and kinematics of
stellar populations) with observations of local galaxies. Finally,
with such models predictions can be made for dynamical and stellar
population properties of high-redshift galaxies, which can be verified
by future high resolution observations with the next generation
telescopes.

The mass accumulation into dark matter halos is well understood in the
context of the cosmological simulations and can be used as an input to
model galactic evolution \citep[e.g.,][]{vdbosch02, wechsler02}.
Additionally, the angular momentum distributions of the forming dark
halos have been calculated recently \citep{bullock01, chen02}.  It is
widely accepted, both in the semi-analytical approach and the
small-scale dynamical models, that galactic evolution depends
sensitively on the mass and angular momentum distribution of the
system.

Much less is known about the processes that govern the evolution of
the baryons within a dark halo, and how these influence the properties
of the forming galaxies.  Stars in galaxies nearby are observed to
form in molecular clouds much denser than the ambient medium in which
these clouds are embedded. Most of the kinetic energy of this cloud
fluid is in the motions of single clouds relative to the bulk flow.
This kinetic energy can be dissipated by inelastic collisions
\citep{larson69} and augmented by supernova feedback \citep{mckee77}.
Under some conditions the macroscopic cloud system can be treated as
an isothermal fluid \citep{cowie80}. Its energy dissipation rate is
not well-determined, however. One expects that it depends on the
geometrical structure of the clouds, on whether a major part of the
dense medium is arranged in filaments, and on their self-gravitating
structure and magnetic fields \citep{kim01, balsara01}. Thus the
dynamics of the macroscopic cloud medium may be more or less
dissipative, depending on the physical conditions, and may well vary
between galaxies.

In the present paper we investigate the formation and dynamical
evolution of galactic disks, varying the cloud dissipation rate.  We
use the interaction network and two-phase chemodynamical model of
\citet[SG03]{samland03} to describe the assembly of a disk of stars,
hot gas, and star-forming cold gas. We find that the dynamical
stability of the disk depends sensitively on the cloud dissipation
efficiency, here described by the parameter $\etac$. For large
$\etac$, dynamical instabilities in the gas dominate the evolution, leading to
fragmentation of the disk into a small number of star-forming clumps
which subsequently merge to form a centrally concentrated bulge. For
small $\etac$, on the other hand, the system forms stars until the
stellar disk becomes unstable, leading to a stellar bar at late times.
The different stability properties in both cases can be quantified in
terms of the effective Toomre $Q$ parameter for the stellar, gaseous,
and combined system \citep[e.g.,][]{toomre64, jog84, wang94,
  elmegreen95}.

The evolution of a system with high dissipation calculated at higher
resolution is described in \citet{immeli03}, where it was shown that
morphological and photometrical properties of several high redshift
objects, like the chain galaxies \citep{cowie95}, can be explained by
a fragmented disk model.

Several authors have suggested that galactic bulges can form by the
secular evolution of a galactic disk, driven by interstellar gas or
stars \citep[e.g.,][]{combes81, pfenniger90, noguchi99}.  Recent observations
lend some support to these secular evolution scenarios.  Bulges in
late-type spirals show similar properties to their surrounding disks, 
in that their light profiles are better fit by an exponential rather
than an $R^{1/4}$ law \citep{courteau96, seigar02}, and in their
colours \citep{peletier96}.

As shown here, the evolution of star-forming galactic disks may take
different routes, depending on whether it is the gas or the stars that
drives a disk instability. This in turn depends on the uncertain
efficiency of energy dissipation in the cold cloud component. These
different routes also lead to different formation scenarios for
galactic bulges.  Thus we suggest here that the morphological
properties of galaxies may depend not only on cosmological variables
like different mass or angular momentum distributions, or infall history, but also on
internal physical processes during galaxy evolution.

In Sect.~\ref{modell} we describe our model for star-forming disks.
In Sect.~\ref{morphologie} we describe the morphological evolution of
these disks, depending on the dissipation efficiency of the cold gas.
Sect.~\ref{globalevo} discusses the stability and star formation rates
in the models, and Sect.~\ref{bulge} describes the properties of the
bulges that form in the two main evolutionary scenarios. Finally,
Sect.~\ref{concl} summarizes our findings.

%
\section{The Model} \label{modell}

We use a two-phase model for the interstellar medium, consisting of a
hot, low-density phase and a cold cloud medium from which stars are
formed. The chemical elements and most of the energy released from
SNeII and later SNeIa are returned to the hot phase. However, the
cloud velocity dispersion of the cold phase is heated by SNeII as well
\citep{mckee77}.  We describe this system with a three-dimensional
chemodynamical evolution code, which combines a hydrodynamical grid
code for the two phases of the interstellar medium (ISM) with a
particle mesh code for the stars. See SG03 for more details.

The interactions between the different ISM phases and the stars are
described in detail in SG03.  A difference exists in the
characterization of the star formation rate (SFR), where we use here a
simple Schmidt law \citep{schmidt59},
\gl{schmidt}{
\rhopsf = \csfr \cdot \rhoc^{\alpha}
}
with $\alpha=1.5$ and $\csfr$ chosen to be consistent with the star
formation rule derived by \citet{kennicutt98}. We here adopt a volume
density star formation threshold $\rho_{\rm{th}}$, converted from the
surface density threshold of $1-10~\msun pc^{-2}$ obtained from
observations by \citet{kennicutt98}, assuming a disk scale height of
the order of the spatial resolution of our model.  Thus
$\rho_{\rm{th}} \simeq 1~\msun pc^{-2} / 500~pc = 0.002~\msun
pc^{-3}$.  This is a lower value compared to other thresholds used in
the literature \citep[e.g. ][]{noguchi99}. It ensures that gas does
not clump simply because it is below a high star formation threshold
and cools when there is no heating from star formation.

SG03 did an extensive investigation of the network of processes
connecting the different ISM phases in this model. We use their values
for most of the efficiency parameters that occur in this network, as
determined through either theory or observation.  As shown by SG03,
this network is strongly self-regulating, so that the system is not
sensitive to the precise values used for these efficiencies.  They
conclude that the cloud dissipation efficiency $\etac$ is the most
uncertain parameter in the model. We now describe the relevance of
this parameter in somewhat more detail.


The dynamics of the cold cloud medium (CM) is described in a
statistical way through the hydrodynamic moment equations. Energy
dissipation and gain are implemented through sink and source terms to
the moment equations.  The internal energy describes the kinetic
energy of the single clouds relative to the bulk motion. The CM cools
through reducing this kinetic energy, which in this model is assumed
to happen through inelastic collisions between single clouds.
\citet{larson69} derived the dissipation rate in a CM consisting of
spherically symmetric clouds with constant mass. Using the mass-radius
relation for clouds given by \citet{elmegreen89} the internal energy
change can be written as
  \gl{larson}{
  \dd{\ek}{t}
    = - \etac
    P_4^{-1/2} \rho^2 \sigma^3
  }
where $\rho$ is the density of the CM, $\sigma$ its velocity
dispersion, and $P_4 = (P_{\rm{icm}}/k)/{10^4}$ with $P_{\rm{icm}}$
denoting the pressure of the intercloud medium .  All constants have
been merged to $\etac$. $\etac$ also contains the deviation of the
effective cross section for cloud-cloud collisions from the
geometrical value due to magnetic fields, self gravity, and
gravitational focusing.  It also depends on whether the
cold cloud medium is arranged mostly in clumpy structures or
mostly in filaments, which is not well-understood. Thus
the precise macroscopic description of cloud dissipation on
a galactic scale is uncertain, and hence the value of
$\etac$ in eq.~\ref{larson}.  It is therefore possible that the 
dissipation efficiency changes from one galaxy to another, 
and that this changes galactic evolution. It is equally possible
that interstellar cloud physics is nearly universal, and that the main
driver for galactic evolution is the baryonic infall rate per unit
area, or disk growth time, as \citet{noguchi99} has argued. In both
cases, the local heating-collisional cooling equilibrium of the cloud
medium is changed, and hence the star formation history.

To investigate the influence of cloud dissipation on disk evolution,
we performed a sequence of simulations with different dissipation
efficiencies $\etac$. The values given in Table~\ref{etatab} are
normalized to the efficiency $\etac^D$ used in SG03. The range of
$\etac$ values corresponds to the maximum variations given in 
Table 1 of SG03.
\begin{table}[h]
\begin{center}
\begin{tabular}{cc|cc } \hline \hline
model index & $\etac$ [$\etac^D$] &  model index & $\etac$ [$\etac^D$] \\ \hline
          A &    20               &            E & 1/5  \\
          B &    10               &            F & 1/10 \\
          C &     5               &            G & 1/20 \\
          D &     1               &              &      \\
\hline
\end{tabular}
\caption{Dissipation efficiencies used in the model sequence,
         normalized to model D, where $\etac^D = 0.025$ (SG03).}
\label{etatab}
\end{center}
\end{table}

The dynamical set-up of the models describes the rapid
formation of a massive galactic disk in a pre-existing, static dark matter
halo with NFW-profile \citep{navarro97}.  
According to \citet{sommer02} the delayed infall
of the baryonic matter into the relaxed halo can solve the angular
momentum problem arising from $\Lambda$CDM structure formation
simulations.  
The primordial gas enters the simulation volume
vertically at $|z|=$ 15.5~kpc, and the infall is uniformly distributed
over a radius of 17~kpc, with a rotation velocity equal to the
circular velocity at the infall point.  The infall rate is $120~\msun
yr^{-1}$ during one Gyr, resulting in a total mass of $1.2 \cdot
10^{11}~\msun$.  The simulation volume has a diameter of 38~kpc and a
vertical height of 31~kpc with a spatial resolution around 500~pc.
The evolution of a higher-resolution simulation of model A is
discussed in \citet{immeli03}.  The higher resolution does not lead
to different results, which indicates that the outcome of our
simulations is not sensitive to the resolution used.

The chemodynamical model naturally provides ages and metallicities of
all stars formed over time, as well as ISM densities and
metallicities.  This enables us to calculate HST- and
UBVRIJHKLM-colours of the model, including dust absorption. This is
done using the method described in \citet{westera02}, but we adopt
here a three times lower absorption coefficient.  We will use the
observable colours to discuss the morphological evolution.

%
\section{Morphological Evolution} \label{morphologie}

%
\subsection{Settling of the Disk} \label{settling}

\begin{figure*}[htp]
        \vspace*{3mm}
        \flushright
        \includegraphics[angle=0,width=.98\linewidth]{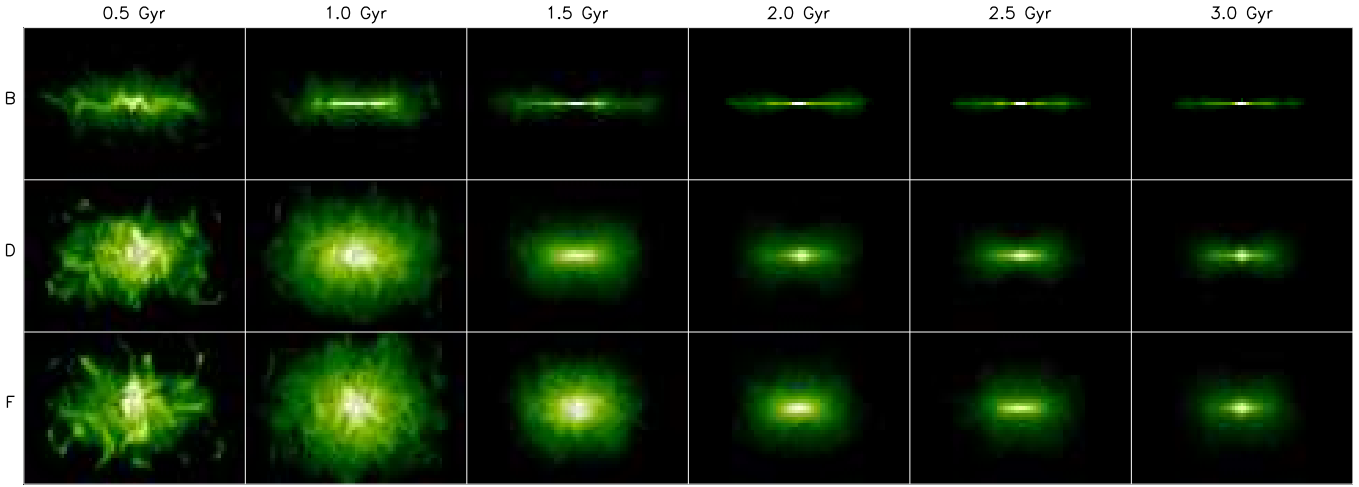}
        \caption{Cut through xz-plane of the cold gas density distribution
             for models B, D, F.
             Evolutionary time and model index are indicated.}
        \label{settlingbild}
        \end{figure*}

The infall of the baryonic matter leads to the build-up of a gaseous
disk.  The settling of the disk is shown in Fig.~\ref{settlingbild},
which shows a cut through the $xz$-plane of the density distribution
for the cold gas component in models B, D and F at different times.
The energy feedback from the first stars heats up the disk. In
consequence the settling of the disk takes longer than a free-fall
timescale, depending strongly on the cloud velocity dispersion and
hence on the cloud dissipation efficiency $\etac$. Since the angular
momentum does not play a role in the settling along the rotation
axis, the $z$-component of the velocity dispersion alone prevents the
clouds from completely falling to the disk plane.

After 500~Myr the cold gas phase shows a filamentary distribution in
all three models. This is caused by the local energy feedback of the
first supernovae of type II (SNeII) to the gas. This leads to local
variations in the velocity dispersion of the cloud medium, which act
like the analogues of pressure gradients in the hot gas phase. Thus
the cold gas is compressed. In addition, in regions with enhanced
energy input, the equilibrium between hot and cold gas phases is shifted
to decrease the density of the cold medium.

The model with high dissipation (B, first row) settles to a thin
disk which is almost completely supported by rotation. The timescale
for the settling is relatively short because the energy input from SNeII
can be dissipated efficiently.
The less efficient energy dissipation of model D causes the disk to
settle more slowly. The disk in this model also remains thicker during
the whole evolution, because the equilibrium between the continuous
energy input from SNeII and energy dissipation by cloud collisions is
shifted to higher internal energies and thus higher cloud
velocity dispersions.
This is even more true in model F, where the even less efficient
energy dissipation results in a yet thicker disk.

%
\subsection{Face-On Evolution}

Fig.~\ref{uebersicht} shows an overview of the face-on morphological
evolution of all our models, presented in terms of the V-band
restframe surface brightness calculated from the distribution of stars
as described in Section 2. Model indices are indicated at the left for
each row, and time is indicated at the top for each column.  All
images are normalized to the same surface brightness magnitude
interval, which enables us to directly compare luminosities of the
models.  The absolute V-magnitude $M_V$ of the galaxy is also given in
each panel. Obviously, the morphological evolution depends strongly on
the dissipation efficiency.
        \begin{figure*}[htp]
        \vspace*{0.5cm}
        \begin{center}
        \includegraphics[angle=0,width=.87\linewidth]{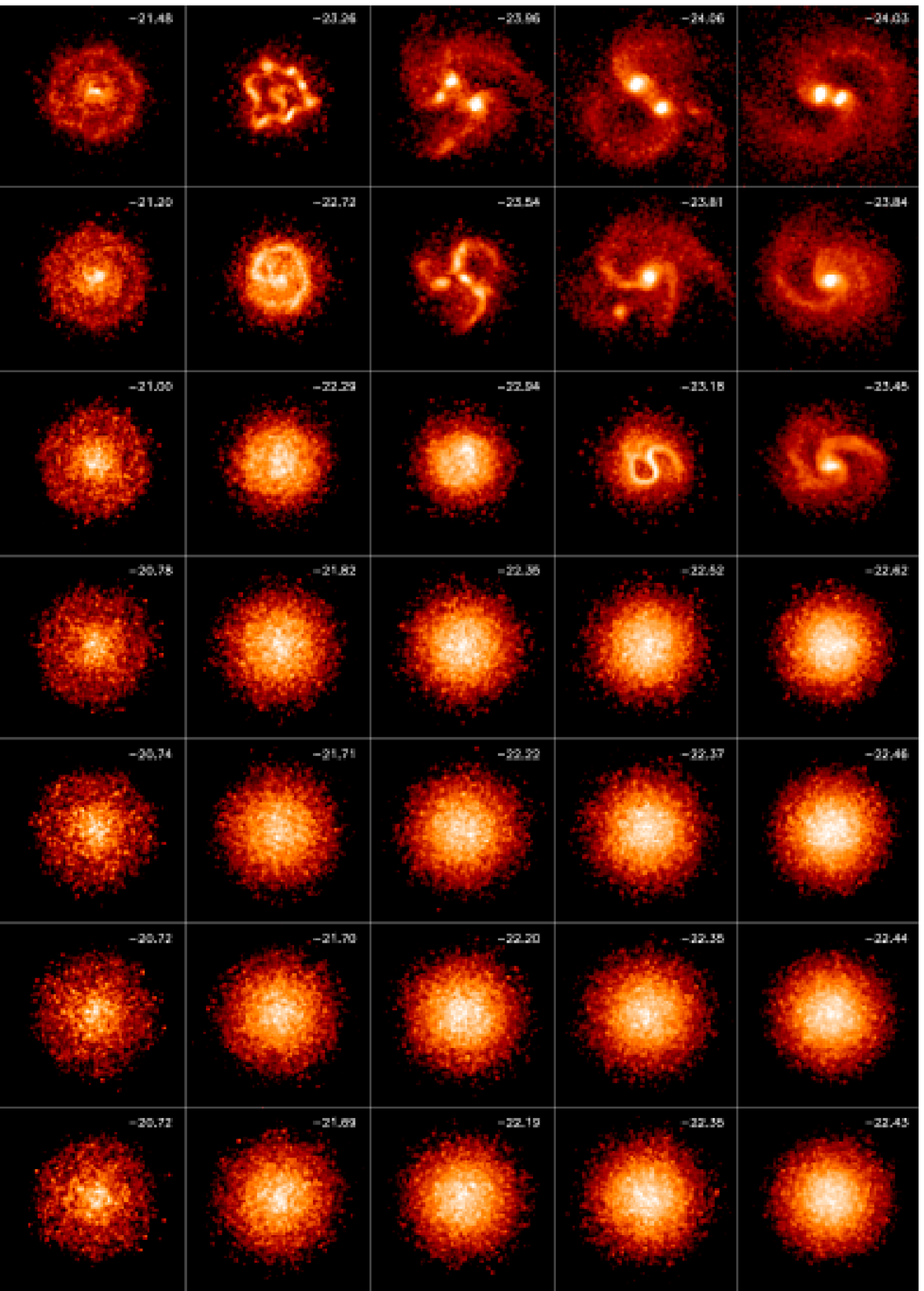}
        \end{center}
        \caption{Morphological evolution of the seven models in terms
                 of V-band restframe surface brightness.  The panels
                 show the whole simulation area of 38~kpc squared. The
                 model indices are given at the left and time is
                 indicated at the top. Between panels, time does not
                 progress in constant
                 intervals; this is to emphasize the interesting evolutionary
                 phases of the models. $M_V$ is given in each panel.}
        \label{uebersicht}
        \end{figure*}
        \addtocounter{figure}{-1}
        \begin{figure*}[htp]
        \vspace*{0.5cm}
        \begin{center}
        \includegraphics[angle=0,width=.87\linewidth]{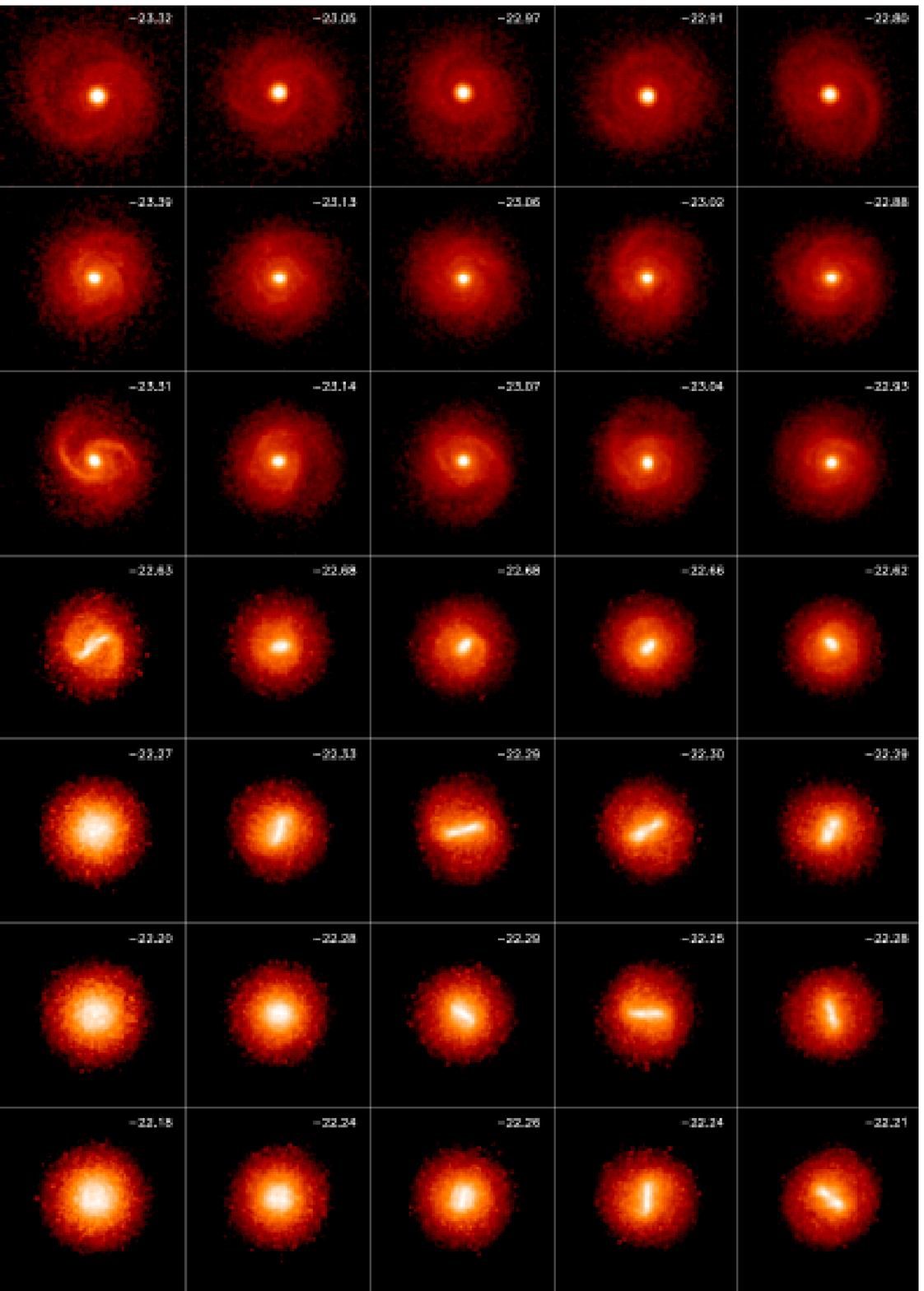}
        \end{center}
        \caption{- {\sl continued}.}
        \end{figure*}

At 0.5~Gyr the formation of the disk is under
way and none of the models shows an asymmetry in surface
brightness yet. Generally, the colder models are brighter in absolute
V-magnitude than their warmer counterparts.
At 0.75~Gyr the coldest model (A) shows several knots of high density
with enhanced star formation. In model B a ring like structure
is visible. 200~Myr later the disks of these two models have fragmented,
showing several nearby knots.
At 1.05~Gyr some clumps in models A and B have already merged and
fragmentation is also visible in model C.
Whereas at 1.2~Gyr models D to G still show a symmetric surface brightness
distribution, the clumps in models A to C have already fallen
to the center and merged to form a bright bulge. This persists for
the remainder of the simulation, together
with a clear spiral pattern in the disk.
Model D develops a bar like structure with spiral arms at around 2~Gyr.
A very pronounced bar is visible in models E to G at around 2.8~Gyr.
As can be seen in the last column of Fig.~\ref{uebersicht}, these bars
become shorter. Bar evolution is discussed further in Sect.~\ref{barevol}.


The morphological evolution of models A, B, and C is similar in
the sense that they all develop a fragmented disk and end up with a
compact central concentration.  Their exists a dependence on $\etac$,
in such a way that the more efficiently the gas dissipates energy, the
more pronounced is the clumping, because the lower velocity dispersion
in the more dissipative models counterbalances the gravitational
instabilities less well.
In all the cold simulations a ring develops. This ring structure
probably results from the form of the rotation curve in these models;
see \citet{cha03, gingold83}.  It is important to note that the mass
enhancement in the ring amounts to less than 10\% of the total mass in the
cloudy medium.

The galaxy models A-C in their fragmented phase resemble, when viewed
edge-on, the chain galaxies first reported by \citet[CHS95]{cowie95}.
These are high redshift galaxies observed with HST in the Hawaii
Survey Fields, with large major-to-minor axis ratios, knotty
structures and very blue colors.  CHS95 suggest that chain galaxies in
the redshift range $0.5-3$ have a mass comparable to that of a
present-day galaxy and that they represent a new population of
galaxies.  Our model favors the interpretation of \citet{oneil00} that
the chain galaxies are clumpy disk-like structures seen edge-on.
Model A, recalculated with higher resolution, is compared to
observations in detail in \citet{immeli03}. The main results are that
the extraordinary morphology as well as very blue colors observed by
CHS95 are reproduced well.  An investigation of the properties of the
individual clumps in this model leads to typical stellar clump masses
of a few $10^9~\msun$.  Metallicity differences between individual
clumps are not larger than 0.25~dex and the influence of individual
clumps on the rotation curve can by as high as $100~{\rm km/s}$.


A similarity in the evolution can also be seen in  models E to
G. In these models the disk remains symmetric for a much longer
time. At around 2.7~Gyr these models then become bar unstable, and a
very prominent bar develops in about one dynamical
time. Contrary to the fragmented disk models, the absolute luminosity
of the models showing a bar instability does not vary much over the
simulated time interval.  The evolution of model D lies somewhere
in between these two cases, showing a bar with spiral arms at around
2~Gyr.

Comparing models A and B shows that no qualitative differences in the
evolution are visible. The same is true for models F and G, which
indicates that the chosen values of $\etac$ (Table~\ref{etatab})
covers the whole relevant range.

%
\section{Global Properties} \label{globalevo}

%
\subsection{Stability of the Disk} \label{diskstab}


As shown in the previous section, our models can be divided in two
groups according to their dynamical evolution: Models A to C show an
early fragmentation, whereas in models E to G a bar instability
occurs at later times. Model D appears to be a transition case. In
the remainder of this paper, we confine the discussion to models
B, D, F, as representative models for the different evolutionary
paths.
First, we relate these more quantitatively to the stability of the
multicomponent disk. The Toomre parameter $Q$
\citep{safranov60,toomre64} is an important quantity in investigations
of disk stability. In a single component gaseous disk $Q$ is given
through
  \gl{q}{
  \Qgas = \frac{ \kappa\sigma_{\rm{gas}} }{ \pi G\Sigma_{\rm{gas}} }
  }
where $\Sigma_{\rm{gas}}$ is the surface density, $\kappa$ the
epyciclic frequency, $G$ the gravitational constant, and
$\sigma_{\rm{gas}}$ is the radial velocity dispersion of the gas.
$Q=1$ is a well-defined stability limit for such disks:
disks with low velocity dispersions and high surface densities such
that $Q<1$ are subject to radial instabilities.


In a disk consisting of gas and stars the evaluation of $Q$ is more
difficult.  As described above, the interstellar material in our model
consists of two gas phases.  Because the hot intercloud medium (ICM)
only makes up for about 1\% of the total gas mass and because it is
pressure dominated and therefore stable in a Toomre sense, we can
neglect the ICM contribution to the Toomre parameter. Henceforth, we
consider a composite $\Qeff$ for the stellar and the cold gas phases.


There have been several attempts to describe the stability of a
two-component systems in the context of the Toomre parameter.
\citet{jog84} derived a dispersion relation for a two-component system
of stars and gas and evaluated it numerically.  \citet{romeo92}
derived finite thickness corrections for disk systems.  A simple
approximation was inferred by \citet{wang94}
  \gl{qeff}{
  \Qeff \simeq \frac{\kappa}{\pi G}
       \left( \frac{\Sigma_{\rm{gas}  }}{\sigma_{\rm{gas}  }} +
              \frac{\Sigma_{\rm{stars}}}{\sigma_{\rm{stars}}}   \right)^{-1}
    \simeq \left(\frac{1}{\Qgas} + \frac{1}{\Qstr} \right)^{-1}
  }
where $\Sigma_{\rm{stars}}$ is the stellar surface density and
$\sigma_{\rm{stars}}$ is the radial velocity dispersion of the stars.
\citet{elmegreen95} gives a more accurate way of calculating the
effective $\Qeff$ parameter for a two component disk. He reformulates
the dispersion relation of \citet{jog84} in such a way that the
derived $\Qeff$ is completely analogous to the one-phase $Q$.
We use this more accurate method as well as the approach of
\citet{wang94} to evaluate $\Qeff$ for the two-phase medium. Since the
results are the same, we will not further distinguish between the two
methods.

\begin{figure*}[htpb] \centering
        \vspace{1.2cm}
        \includegraphics[angle=0,width=.95\linewidth]{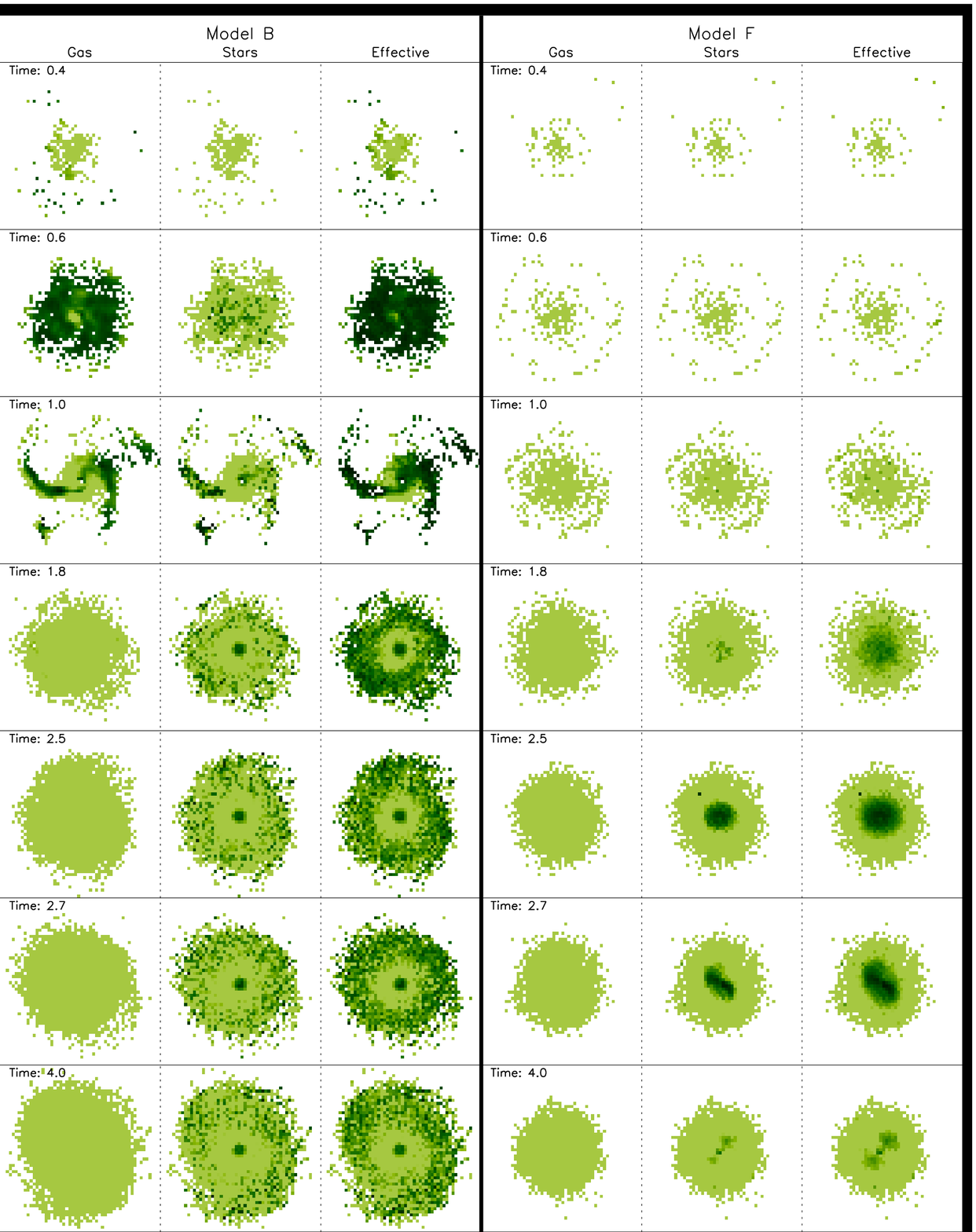}
        \vspace{0.5cm}
        \caption{Toomre instability map for model B (left panels) and
          model F (right panels).  The $Q$ Parameter is given for the
          cold gas component and for the stars, and the effective $Q$ is
          shown for the two-component system. Grey (light green), dark
          (dark green) and white colour in this map denotes regions of
          stability, instability, and undefined stellar and effective $Q$.
          Time is indicated in
          Gyr. The panels have a side length of 25~kpc.}
        \label{toomremap}
        \end{figure*}

Fig.~\ref{toomremap} shows maps of $\Qgas$, $\Qstr$, and two-component
$\Qeff$ for models B and F. The white regions in the maps denote those
parts of the disk where the number of stellar particles in a grid cell
in the plane is less than five and no secure stellar velocity
dispersion and hence no reliable $\Qstr$ could be determined. These
maps show clearly that the $Q$ values can vary substantially between
different parts of the disk, both radially and azimuthally.

At around 0.4~Gyr, not many cells reach the stability boundary
just discussed: all $Q$'s have values above unity.
Already at 0.6~Gyr the situation changes in model B. The gaseous component
develops regions of instability, and $\Qeff$ reveals an unstable
disk over a large radial range.
For comparison, the disk of model F is still stable at 1~Gyr in both
components as well as in the composite $\Qeff$. In model B the violent
evolution following the instability causes a sharp increase in the
mean $\Qeff$. The instability converts energy from ordered motions to
random motions, as well as driving up the SFR and SN heating, which
indirectly increases the cloud velocity dispersion and hence the
Toomre parameter.

At 1.8~Gyr the stellar disk of model F has a central Toomre parameter
around 1.5. The gaseous phase has a Toomre parameter of at least 2 and
therefore is stable not only to radial but also to bar instabilities
\citep{polyachenko97}. Nevertheless, the composite $\Qeff$ already
approaches values around 1 in the center. Thereafter, $\Qeff$ drops further in
the center, causing a bar instability at around 2.6~Gyr.  The bar is
visible in the instability map, but the signature becomes weaker as
the bar evolves.

Globally, both models are stable after their respective instability,
indicating that an equilibrium state has been reached and no further
large morphological changes will occur (compare Sect.~\ref{barevol}.
However, model B remains marginally unstable in its outer parts, which
drives the spiral pattern visible at late times
(Fig.\ref{uebersicht}).  Consistent with the fact that this model has
a thinner disk with higher rotational support than model F, the outer
regions of the disk in model B also have a lower $\Qeff$ than in model
F.

In summary, the most striking difference between the two models is
that the instability in model F occurs in the stellar disk in the
central regions, at comparably late times, whereas in model B the
instability develops in the cold gas phase over the whole of the disk,
starting early-on.  Thus the different morphological evolution
discussed in the previous section can be attributed to the instability
emerging in different phases.

\begin{figure}[h] \centering
        \includegraphics[angle=0,width=\linewidth]{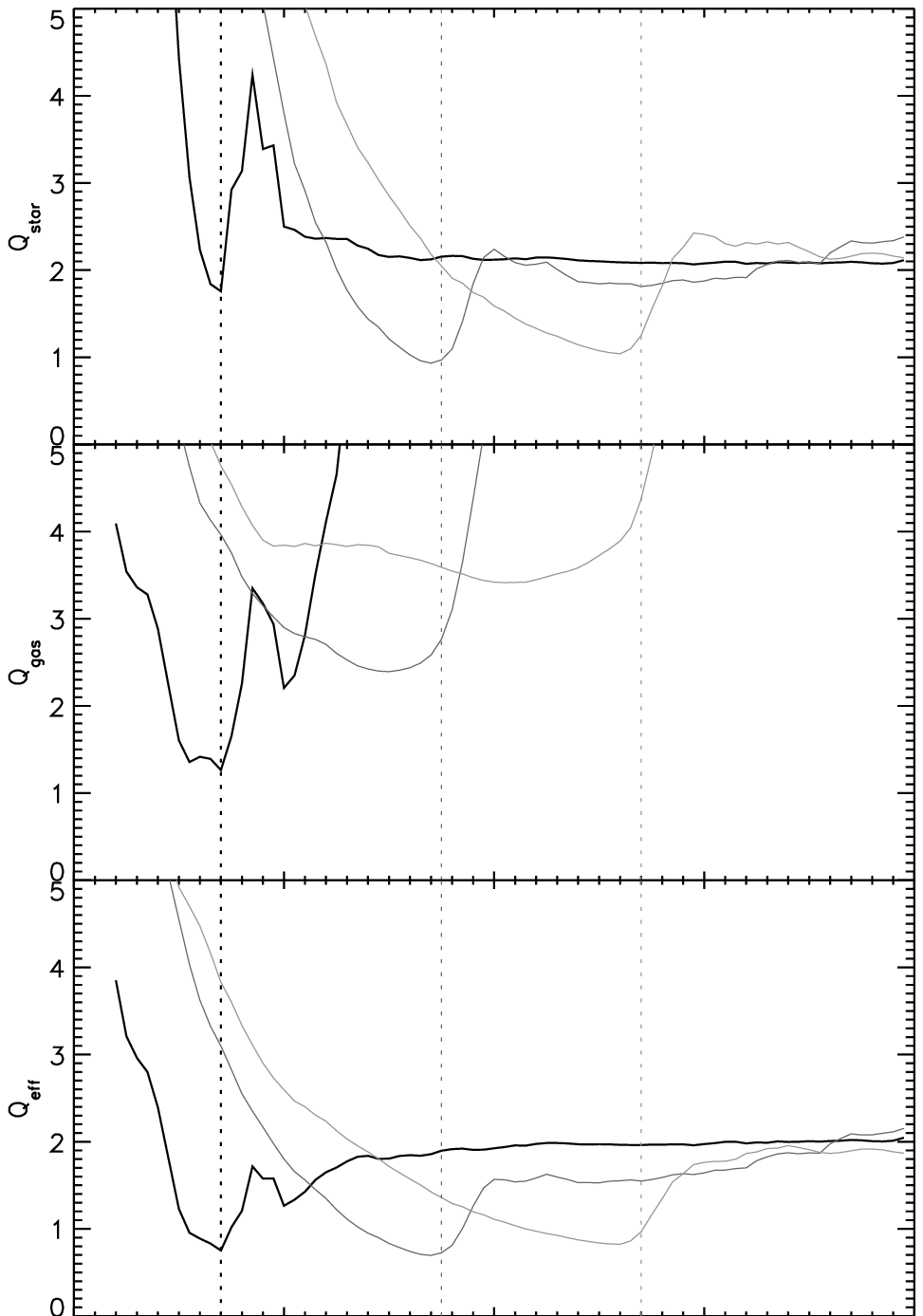}
        \vspace{0.5cm}
        \caption{Time evolution of the Toomre parameter for the
                 stellar component (top), the cold gas component
                 (middle), and the composite Toomre parameter (bottom),
                 for models B (black), D (dark grey) and F (light
                 grey).  The locations where the instability emerges
                 are indicated by dashed lines.}
        \label{toomre}
        \end{figure}

To investigate the time evolution of the Toomre parameter, we averaged
it over a central circle with radius 2~kpc. The progression of this
average $Q$ is shown in Fig.~\ref{toomre}. Again we see that different
components trigger the instability in the different models.  In all
models, the instability starts when $\Qeff \simeq 0.8$. This value
should not be taken as an absolute limit for stability, however, since
it depends on the radius over which the Toomre parameter is averaged.
Nevertheless, this shows that $\Qeff$ is a good tracer of the instability.
Also clearly visible is the increase of $Q$ after the onset of
instability.

%
\subsection{Star Formation Rate} \label{sfrsection}

\begin{figure}[h] \centering
        \includegraphics[angle=0,width=\linewidth]{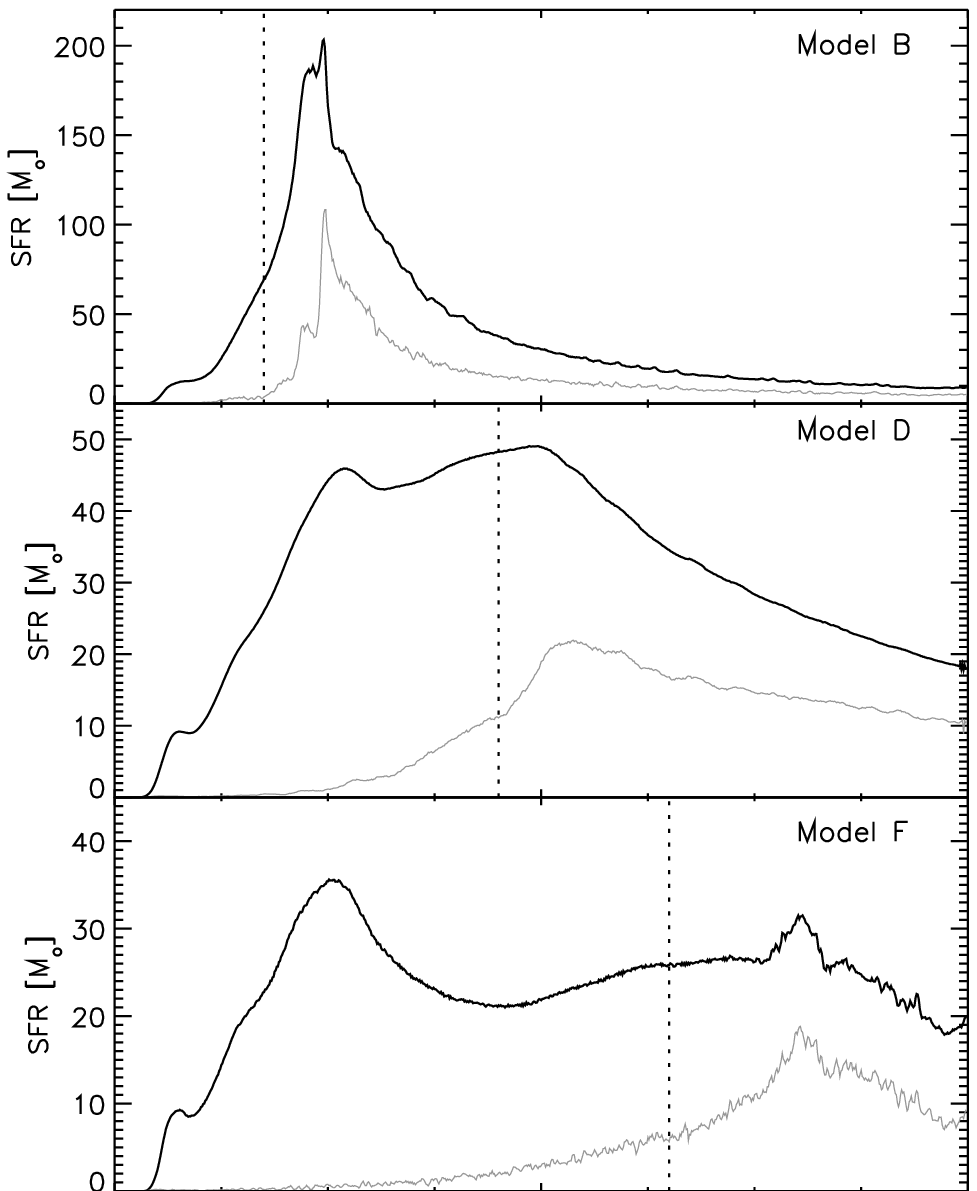}
        \vspace{0.5cm}
        \caption{Global SFR (black line) and SFR in the innermost 2~kpc 
                 (grey line) in models B, D, and F. In all three models 
                 the instability, indicated by the respective dotted line, 
                 causes an increase in the central SFR.}
        \label{sfrs}
        \end{figure}

The different dynamical evolution along the sequence of models has a
strong impact on the global star formation history.  Fig.~\ref{sfrs}
shows the SFR as a function of time, for models B, D, and F. One
immmediately recognizes large differences in the absolute values as
well as in the shape of the SFR.
In the beginning, the SFR is low because the gas must first fall to the
Galactic plane and reach the limiting volume density for star
formation, $\rho_{\rm{th}}$.  This takes around 200~Myr. After this the
SFR increases as more material falls in.  The energy feedback from the
first stars heats the gas and the equilibrium found depends on the
energy dissipation rate of the cold gas phase.
The less efficiently energy can by dissipated, the higher the cold
cloud velocity dispersion, and hence the more extended in the
$z$-direction the cold gas disk is (Fig.~\ref{settlingbild}).  
If, on the other hand, the dissipation efficiency is high, the
velocity dispersion drops and the material settles in a high density
disk. Since all models have identical mass infall and total mass, the
density in the disk plane is lower in the thicker disks. Because the
SFR follows a Schmidt law (Eq.~\ref{schmidt}), the overall SFR will be
higher in the models with strong gas dissipation. This effect can be
seen already at around 500~Myr, when the SFR rates of the models
differ by a factor of two.

The subsequent shape of the SFR is dictated by the further evolution.
In models A to C the cold gas phase of the disk becomes unstable, with
the stars following the potential perturbations induced by the gas.
The clumping of the gas leads to a strong enhancement of the SFR
(Fig.~\ref{sfrs}, top panel).  The clumps then fall to the center and,
due to the efficient angular momentum redistribution during this
phase, a large fraction of the gas is located in the center after the
clumps have merged. In this stage, the SFR reaches a pronounced
maximum, which can be as high as $200~\msun yr^{-1}$.

The warmer models form stars at a much lower rate, after the initial
rise driven by the infall.  Because the gas phase remains stable, the
cold cloud disk from which the stars form has no large density
enhancements. Star formation then increases again when the instability
of the stellar disk sets in. The bar causes an inflow of gas to the
center, causing an increase in the central SFR which also reflects in
the global SFR (Fig.~\ref{sfrs}, bottom panel). Indeed,
\citet{aguerri99} found a strong correlation between the global SFR
and the barred structure of galaxies.

\begin{figure}[h] \centering
        \includegraphics[angle=0,width=\linewidth]{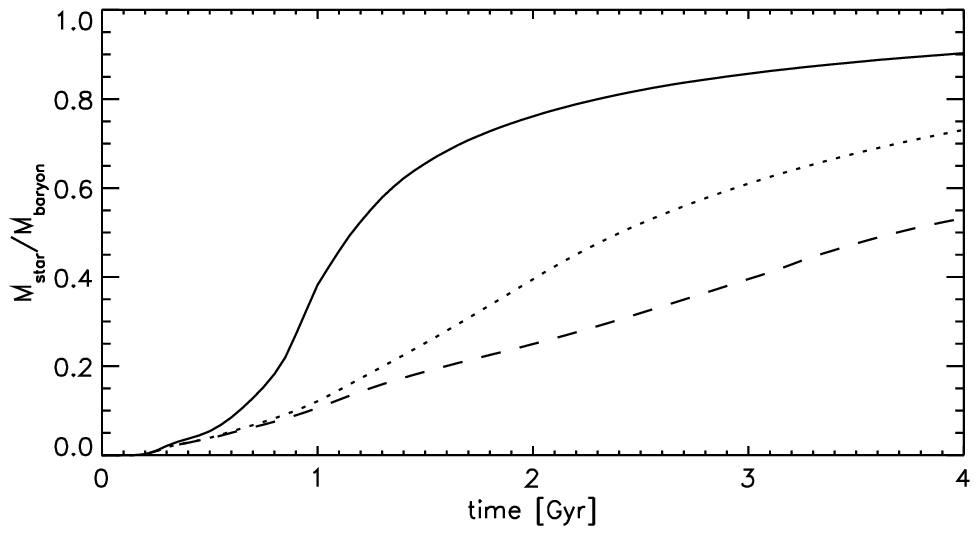}
        \caption{Stellar mass fraction of models B (solid line),
                 D (dotted line) and F (dashed line).}
        \label{massant}
        \end{figure}

As a consequence of their diverse SFR, the integrated mass of stars
formed in these models also varies strongly (Fig.~\ref{massant}):
whereas after 2~Gyr model B has already converted 75\% of its baryonic
mass into stars, the corresponding value for model D is around 40\%
and that for model F is around 25\%.

This illustrates how different the SFR in these models are despite
their identical mass infall rates. The cooling efficiency, acting as a
process on small scales, thus influences the strength and the shape of
the SFR, consistent with results from \citet{vdbosch02}.  It is
therefore difficult to constrain the mass accretion rate of a galaxy
from the star formation history alone, without further assumptions.
This shows the need for self-consistent modelling of galaxies.

%
\subsection{Metallicity Distribution} \label{metdist}

\begin{figure}[h] \flushright
        \vspace{0.5cm}
        \includegraphics[angle=0,width=0.9\linewidth]{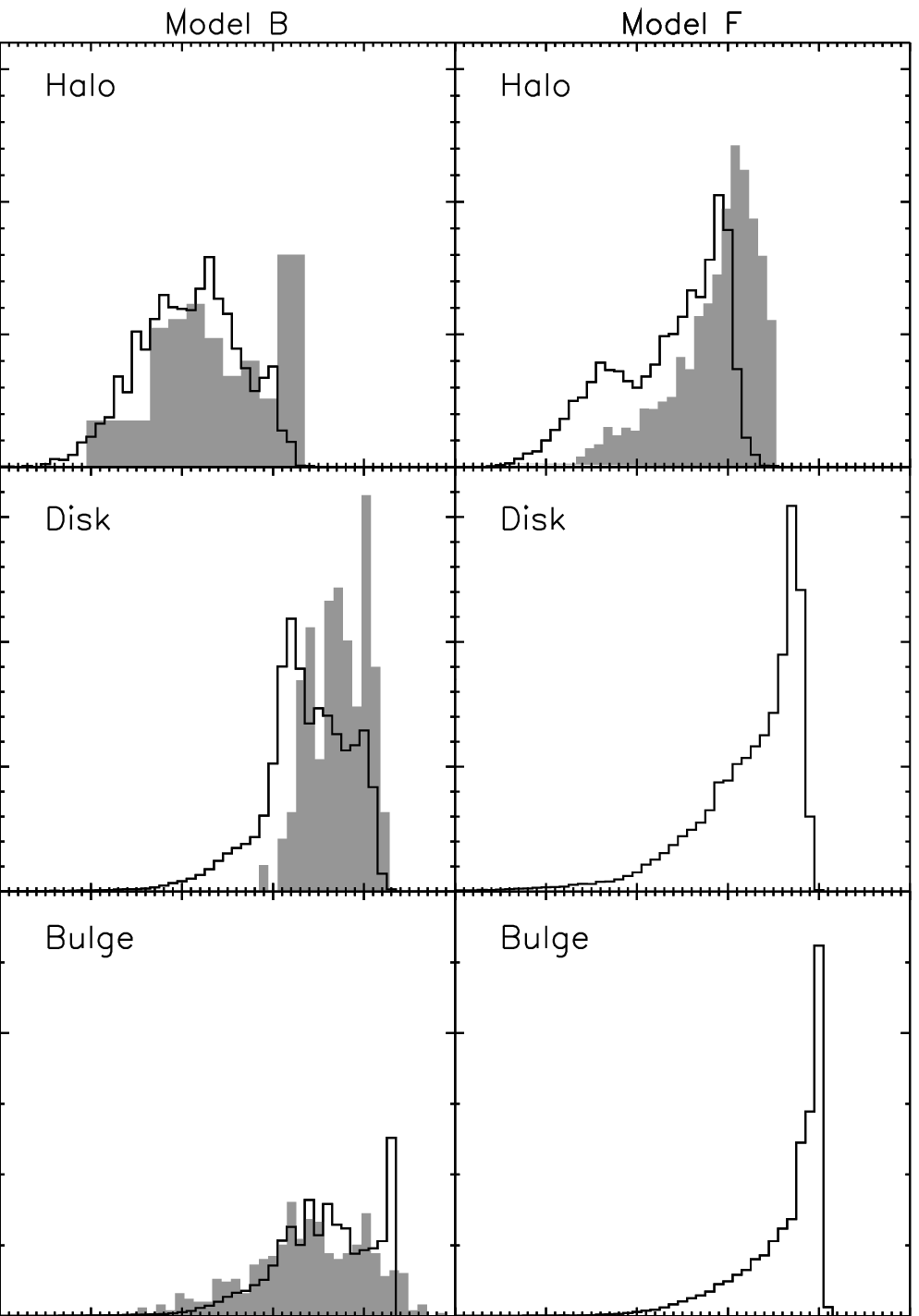}
        \vspace{0.5cm}
        \caption{Distribution of stars in [Fe/H] for models B (left) and
          F (right), for Halo, Disk and Bulge stars as defined in the
          text (open histograms). Observational data are plotted as the
          shaded histograms (references in the text).}
        \label{metvert}
        \end{figure}

Fig.~\ref{metvert} shows the metallicity distribution of the stars in
models B and F after time 4~Gyr.  Here we divide the stars in the model
into halo, disk, and bulge components. The halo is defined as all
stars with a distance to the disk plane of at least 3~kpc. The disk
component is defined as all stars with $z$-distance from the plane
$|r_z| < $ 0.5~kpc and a minimal distance of 2~kpc from the center. All
stars within 2~kpc from the center are taken to be the bulge
component.  

There exist clear differences in the metallicity distributions of both
models, due to their different evolutionary histories. However, in
both models the metallicity increases from the halo to the disk to the
bulge.  The halo has the lowest metallicities because the
metal-enriched clouds falls through the halo towards the disk plane. When
the clouds settle in the disk, the chemical enrichment can take place
over a longer time-scale. The subsequent angular momentum
redistribution from the instability causes the gas to flow inwards.
The inflowing gas cannot escape from the galactic center, and when it
is converted to bulge stars, it has the highest metallicities.

Since the settling of the baryons to the plane depends on the
dissipation rate (Sect.~\ref{settling}), the enrichment time-scale for
stars that end up in the halo is longer for model F than for model B,
and so one expects a more advanced chemical enrichment of the halo in
model F.  Indeed, the halo metallicity distribution of the cold model
B shows a peak at ${\rm [Fe/H]} \simeq -2$, whereas the peak in the
distribution for model F lies approximately one dex higher at ${\rm
[Fe/H]} \simeq -1$.
The shape of the observed metallicity distribution of the halo of M31
\citep{durrell01} is very similar to that of model F. This points to a
relatively slow star formation process for the M31 halo stars.  By
contrast, the metallicity distribution of the Milky Way halo
\citep{chiba00} resembles more the distribution created in the cold
model B, and points to a faster settling of the gas to the disk in the
early collapse of the Milky Way (see also SG03).


In the distribution of disk stars, the differences between both models
are also clearly visible.  In model B, the peak at relatively low
metallicities mirrors the peak in the SFR at early times, before SNeIa
play a significant role in the enrichment.  This model is compared to
data from \citet{edvardsson93}.  Note, however, that the observed
sample contains stars that have formed much later than at 4~Gyr
where the model calculation ends.  The slower star formation process
of model F leads to a steady increase of stars with high
metallicities; in this model the gas is enriched by SNeIa before a
large fraction of the stars is formed.


A similar argument applies to the bulge. Here the high early SFR of
model B produces many metal poor bulge stars with ${\rm [Fe/H]} \simeq
-0.7$.  In model F, by contrast, the evolution is slower, leading to a
steady increase in the stellar metallicity without a peak at low
metallicities.  \citet{tiede99} reported a metallicity distribution for
the Galactic bulge with a similar shape as produced in model B. On the
other hand, results from \citet{ramirez00} and \citet{zoccali03} point
to a more metal rich distribution. The formation of the Galactic bulge
is discussed further in the next section.

Although clearly the cloud energy dissipation is not the only physical
process that influences the metallicity distributions in galaxies, it
does have an important impact on the dynamical, star formation, and
chemical evolution of galaxies.

%
\section{Bulge Formation} \label{bulge}

In our series of models we observe two qualitatively different paths
to bulge formation. In those models where the gas disk fragments and develops
massive clumps, the bulge is formed through the merging of these clumps at
relatively early times \citep[see also][]{noguchi99}.
In the class of models where a bar forms through
an instability of the stellar disk, the bulge is formed from this bar,
at comparably late times. This is similar to the scenario proposed by
\citet{combes81}, \citet{pfenniger90} and \citet{raha91}.

These two routes do not cover the whole range of bulge formation
processes: further ways in which a bulge may be formed are through
mergers \citep{aguerri01, scannapieco03} or in the center of an early
galactic collapse \citep{eggen62}.  It is possible that all these routes
may play a role in the formation of some bulges. Real bulges may even
form through a combination of these processes, as indeed happens in
the self-consistent models of SG03. In particular, some nuclear bulges
would appear to be good candidates for the rapid SFR processes.

The least well-studied of these bulge formation processes is that from
merging clumps in a fragmenting disk \citep{noguchi99}.
\citet{immeli03} compare a high-resolution simulation of model A with
observations and conclude that this process may in fact be at work in
the chain galaxies found by CHS95. Further observations of similar
objects will thus be highly interesting.

\begin{figure}[h] \centering
        \vspace{0.5cm}
        \includegraphics[angle=0,width=\linewidth]{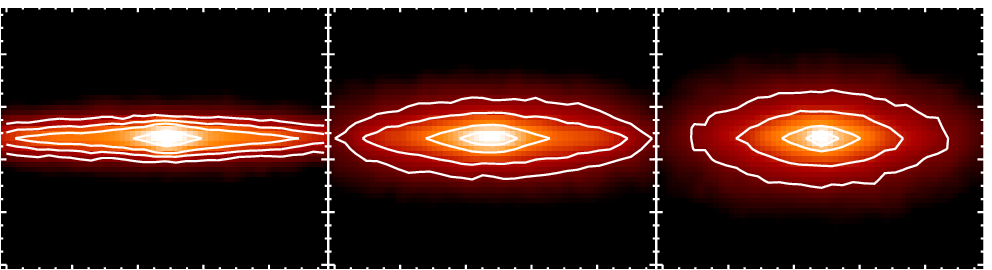}
        \caption{Edge-on view  of the inner 11~kpc of models B, D, and F
          (from left to right), in K-band at time 3.8~Gyr.}
        \label{bulgek}
        \end{figure}

\begin{figure}[h] \flushright
        \includegraphics[angle=0,width=.9\linewidth]{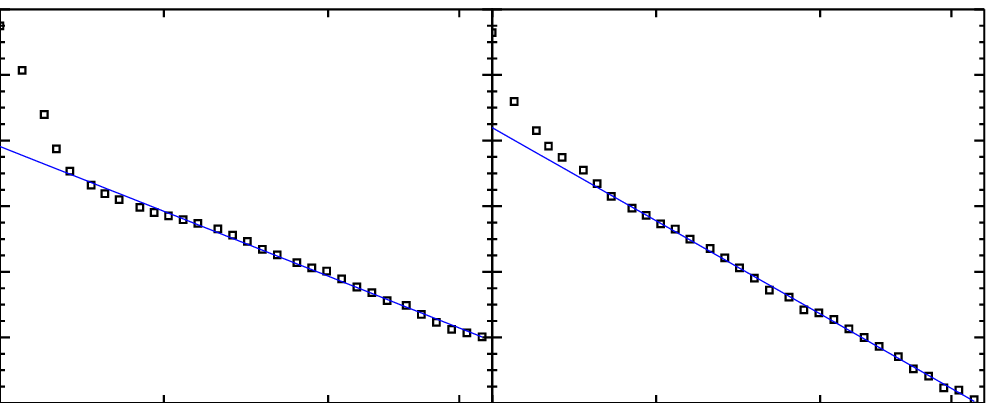}
        \vspace{0.5cm}
        \caption{Radial surface brightness profile in K-band for
          models B (left) and F (right). The violent evolution in
          model B leads to a more massive bulge than the mass transport
          to the center induced by the bar in model F.}
        \label{kprofile}
        \end{figure}

Fig.~\ref{bulgek} shows edge-on projections of the inner regions of
models B, D, and F in terms of K-band surface brightness. The bulges
also stand out in the K-band radial surface brightness profiles,
plotted in Fig.~\ref{kprofile}.  In the following subsections, we give
some more specific predictions from our models.

%
\subsection{Clump Merging} \label{clumpmerge}
\begin{figure}[h] \centering
        \includegraphics[angle=0,width=\linewidth]{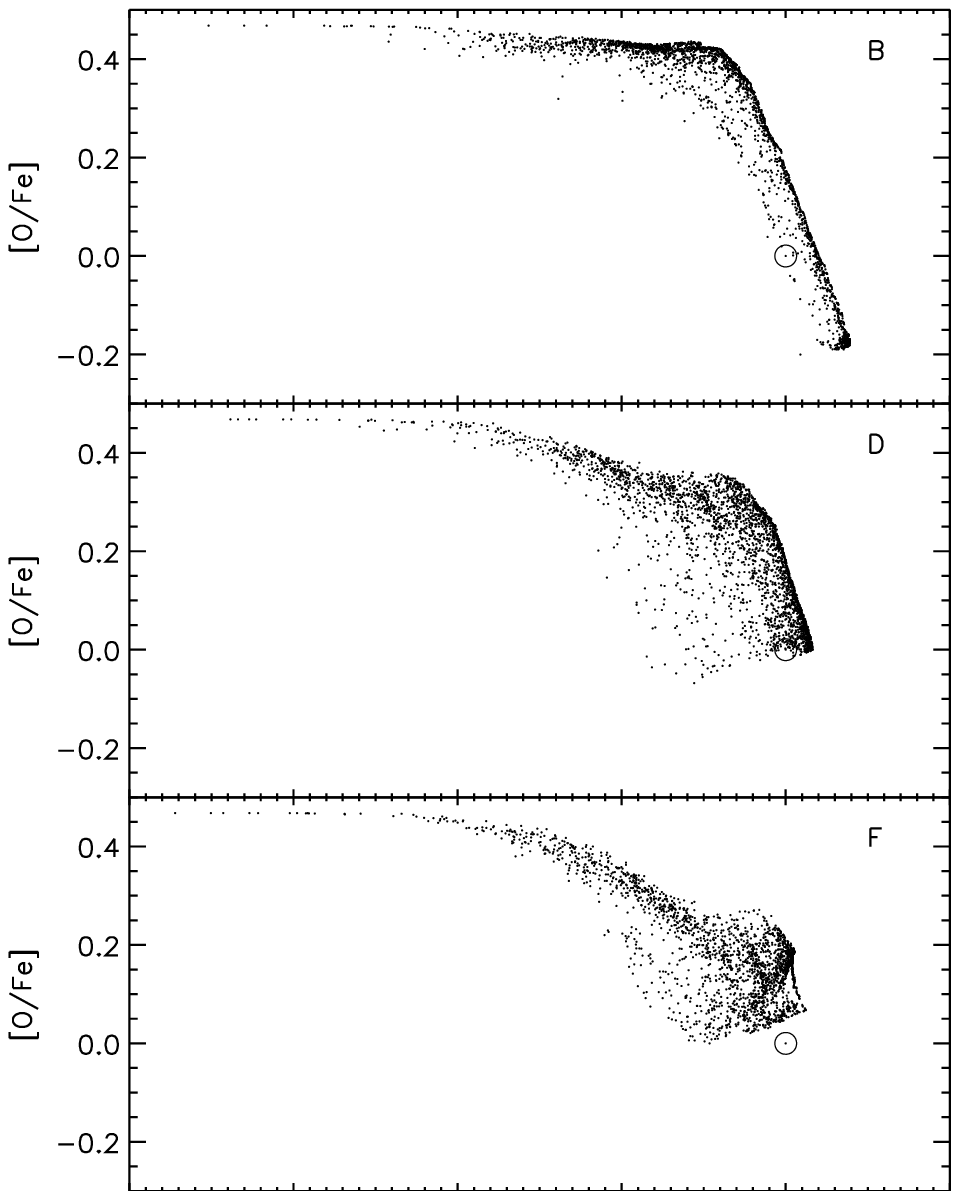}
        \vspace{0.5cm}
        \caption{[O/Fe] against [Fe/H] for bulge stars in the three models 
          B, D, F. The different transitions from SNeII to
           SNeIa dominated abundances are clearly visible.}
        \label{metdistr3}
        \end{figure}

As we have seen, in the cold models (A to C) the disk fragments in an
early evolutionary state and quickly develops several clumps of stars
and gas. The clumps then spiral to the center, where they merge,
causing a massive bulge to form.
Because this process leads to very high gas densities, the SFR in the
clumps and, especially, in the subsequently forming bulge is very
high. This results in a rapid oxygen enrichment of the bulge stars, as
can be seen from Fig.~\ref{metdistr3}, where [O/Fe] is plotted against
[Fe/H] (top panel for model B).

In this plot,  [O/Fe] at low metallicities corresponds to the IMF
averaged abundance ratio of supernovae of type II (SNeII) \citep{samland98},
which enrich the ISM on a timescale of order 10~Myr.
The supernovae of type Ia (SNeIa), assumed in our model to arise from
white dwarf binary systems on a timescale of around 1~Gyr, influence
the chemical evolution of the ISM only at later times.  The large
amount of iron produced by the SNeIa lowers [O/Fe]. The metallicity at
which this happens depends on how far the enrichment has progressed
before this time, and so the location of the transition in
Fig.~\ref{metdistr3} gives information about the star formation
time-scale. In the case of the fragmenting model B, the short
starburst produces a clearly defined path through the [O/Fe]-[Fe/H]
diagrams, with a bend lying around ${\rm [Fe/H]} \simeq -0.4$. Thus
SNeII have enriched the ISM already to this high metallicity before
the feedback from SNeIa plays an important role.

\begin{figure}[h] \centering
        \includegraphics[angle=0,width=\linewidth]{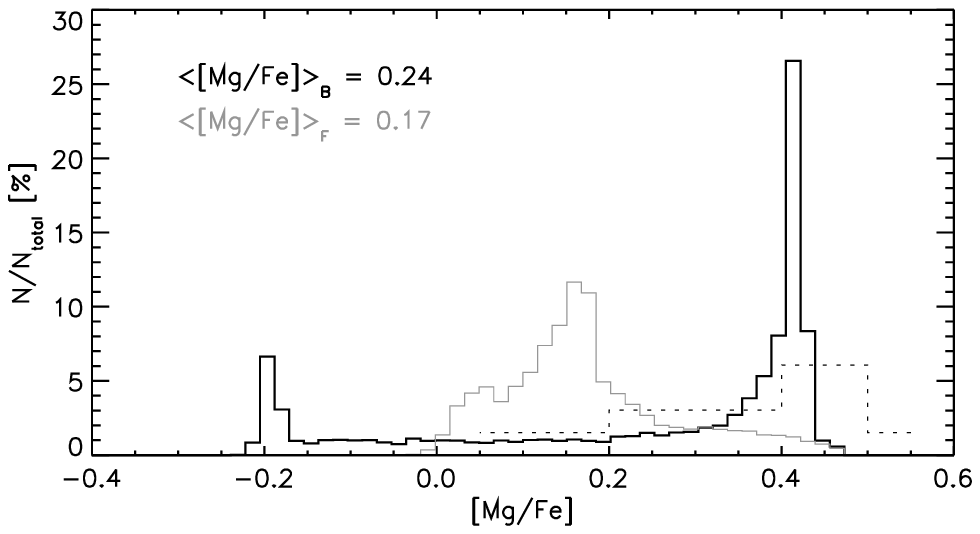}
        \caption{[Mg/Fe] distribution for bulge stars ($r<2\kpc$) formed
          until the end of the simulation (at time 4 Gyr), in models B 
          (black line) and F (grey line). The dotted histogram shows 
          [Mg/Fe] for 11 stars measured by \citet{mcwilliam94}.}
        \label{metdistr4}
        \end{figure}

After the starburst, 70\% of the infalling baryonic matter in model B
has been converted to stars.  The fast SF history is thus reflected in
a large overabundance of $\alpha$-elements. This is shown for the
element Mg in Fig.~\ref{metdistr4}. The distribution of [Mg/Fe] for
all bulge stars at time 4~Gyr shown in Fig.~\ref{metdistr4} has a
strong peak at [Mg/Fe]~$=0.4$.  There is also a smaller peak at ${\rm
  [Mg/Fe]} \simeq -0.2$; this value corresponds to the maximum ${\rm
  [Mg/Fe]}$ reached in a closed box simulation, and is generated
by an extended low level star formation at late times when the iron enrichment
from SNeIa dominates.

The distribution in Fig.~\ref{metdistr4} for model B looks not unlike
that for the Galactic bulge stars measured by \citet{mcwilliam94}
and shown as the dotted histrogram in Fig.~\ref{metdistr4}.
Note, however, that the observed distribution will depend on the
entire SF history following the simulated early phase. Thus, at the
moment it is too early to say whether the old Galactic bulge could
have formed out of an early fragmenting disk, but it is clearly a
possibility that is worth further investigation.

\begin{figure}[h] \centering
        \includegraphics[angle=0,width=\linewidth]{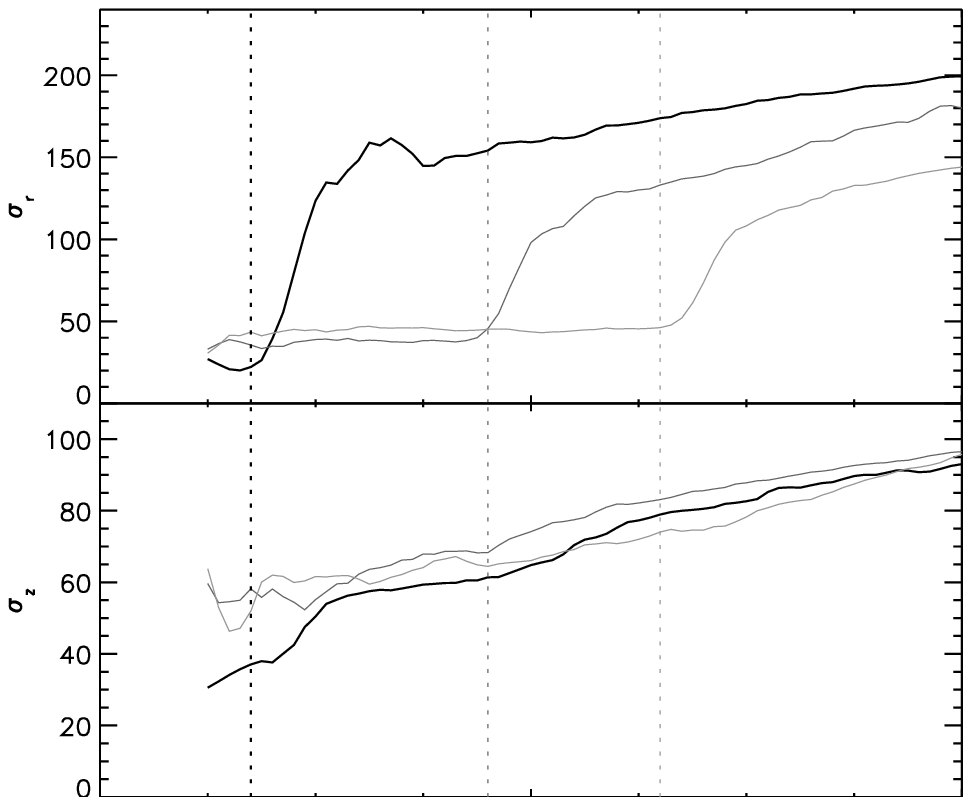}
        \vspace{0.5cm}
        \caption{Stellar velocity dispersion of the bulge stars ($r
                <$~2~kpc) for models B (black), D (dark grey), and
                F (light grey).  The effect of the instability
                on the velocity dispersion
                is clearly visible in the radial component $\sigma_R$.}
        \label{dispbulge}
        \end{figure}

The evolution of the velocity dispersion of the bulge stars is shown
in Fig.~\ref{dispbulge}, where both the (cylindrical) radial velocity
dispersion $\sigma_R$ and the vertical component $\sigma_z$ are given.
The figure shows two main effects:
The first, visible in in both components, is a more or less linear
rise of the velocity dispersion due to the growing mass
concentration in the center of the disk.

The second effect originates in the instability and affects only the
radial component. In model B, the violent evolution of the galactic
disk with the final coalescence to a nuclear bulge leads to a strong
rise in the radial velocity dispersion: when the instability sets in,
$\sigma_R$ rises sharply from around 30~$\kms$ to 150~$\kms$. The
corresponding rise in models D and F occurs later; see Section
\ref{barevol}.  The $z$-component of the velocity dispersion increases
less through the instability.

At early times, the vertical dispersion in this model exceeds
$\sigma_R$.  We believe this is a combination of the infall model used
and a massive disk building up in the galactic plane.  At the end of
the simulation, the radial velocity dispersion is around 200~$\kms$
whereas the $z$-component lies around 90~$\kms$.

%
\subsection{Bar Evolution} \label{barevol}

        \begin{figure*}[thp] \centering
        \includegraphics[angle=0,width=\linewidth]{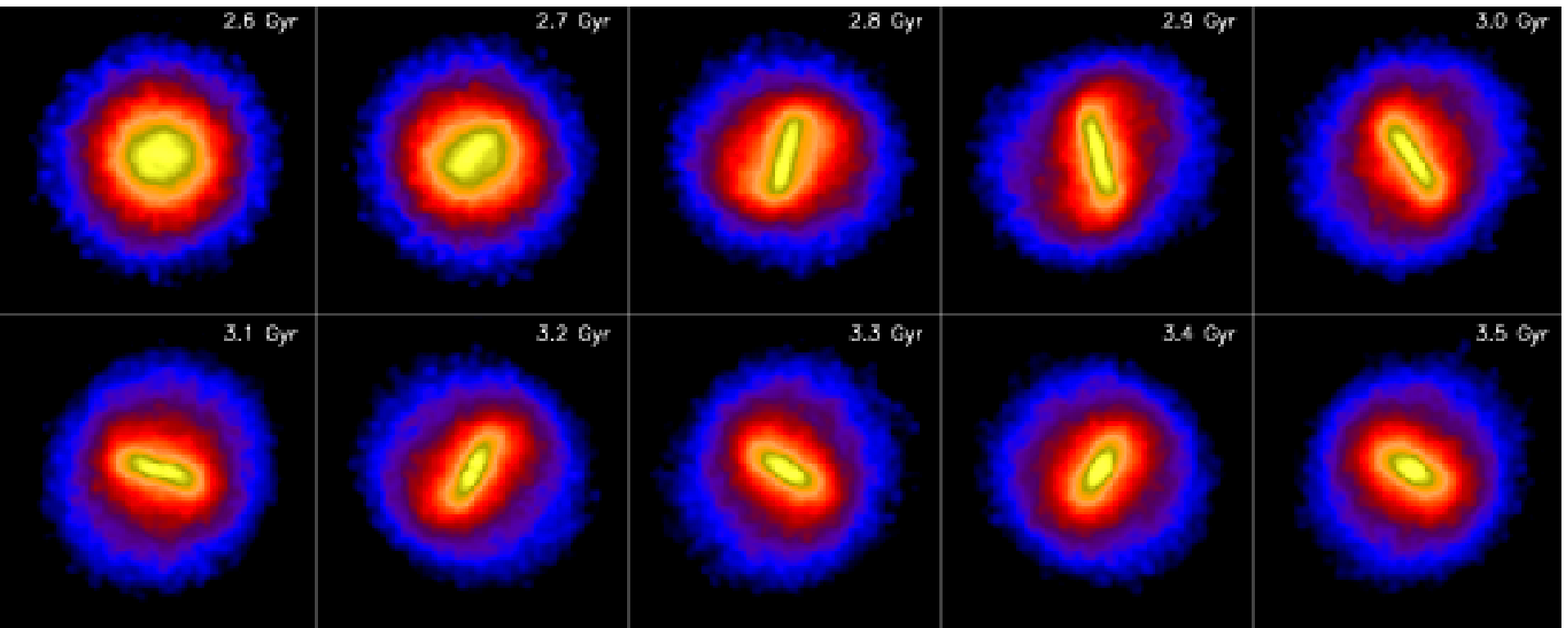}
        \caption{K-band restframe surface brightness map as mass tracer
                 for the inner 12~kpc of model F.}
        \label{barinst}
        \end{figure*}
Fig.~\ref{barinst} shows the evolution of model F during the bar
instability, through its face-on K-band surface brightness map. Model
F is representative for the models with lower cloud energy
dissipation, E to G.
The bar forms between 2.6~Gyr and 2.8~Gyr, i.e., within a time
interval of the order of the dynamical time, as is also observed
in N-body simulations \citep{combes81}.  
The bar instability causes a sudden increase in the radial component of the
velocity dispersion (Fig.~\ref{dispbulge}), whereas the $z$-component
stays rather unaffected.

\begin{figure}[h] \centering
        \includegraphics[angle=0,width=\linewidth]{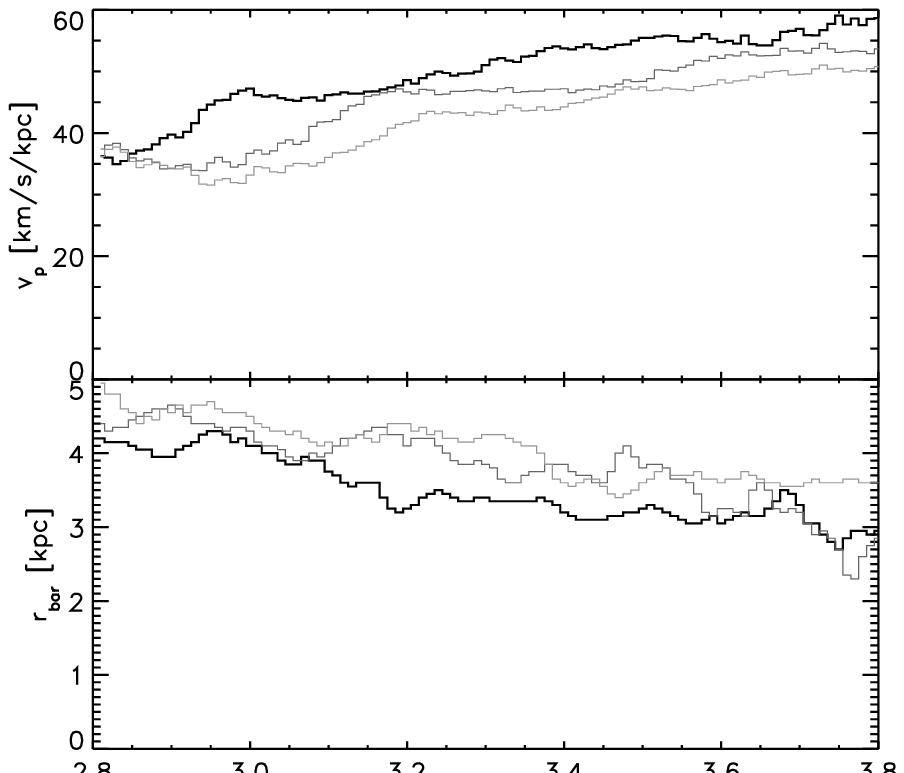}
        \vspace{0.5cm}
        \caption{Evolution of the pattern speed (top) and bar length (bottom)
                 for models E (black), F (dark grey) and G (light grey).}
        \label{pattern}
        \end{figure}

Right after its formation the bar becomes very pronounced, and then
weakens again.  To do a quantitative investigation, we made a Fourier
decomposition of the mass distribution to derive the length of the
bar.
Fig.~\ref{pattern}, bottom panel, shows that the bar becomes shorter
with time. The bar length evolves from around 4.5~kpc to around
3.2~kpc within 1~Gyr. The shortening of the bar is accompanied by an
increasing pattern speed (Fig.~\ref{pattern}, top panel).  The bar
developed in the model is a fast bar with its end near the corotation
radius. The rising pattern speed can be explained by the growing mass
concentration in the center, since all relevant frequencies at a given
radius $r$ depend on $\sqrt{M_{\rm{central}}}$.  This evolution is
similar in all three models that form a bar.

Because at the time when the bar forms (Fig.~\ref{massant}), the gas
content of models E to G is still high, up to 60\%, the formation of
the bar leads to a significant inflow of gas to the center.  There the
gas can form new stars (compare Sect.~\ref{sfrsection}).  The mass
accumulation in the center then weakens the bar, as previously
observed in numerical simulations \citep[e.g][]{friedli93, norman96}.
There is also observational evidence for the weakening of bars through
central concentrations: \citet{das03} recently found an
anti-correlation between central mass concentration and deprojected
bar ellipticity in a sample of 13 barred nearby galaxies.

As we have already mentioned in the introduction, observations
indicate an absence of bars at intermediate and high redshifts
\citep{abraham99}. 
In the models presented here, the formation of an extended stellar
disk with surface density high enough that it can become bar-unstable,
takes a comparably long time.  In addition, in these systems with
large gas fraction, the bar formed is rapidly weakened due to strong
gas transport to the center.  Both effects could occur in high
redshift disks, as the early disks might have larger cloud random
motions because of the stronger infall, and would have higher gas
contents. The two effects thus might explain the absence of bars at
higher redshifts.


The metallicity distribution of the stars in the central bar/bulge
(Fig.~\ref{metdistr3}) reflects the slower enrichment process in the
hotter models. The bend in [O/Fe] is shifted to lower [Fe/H], because
of the smaller SFR in these models.  The bump visible in this plot for
models D and F at [Fe/H] around -0.4 is mainly due to the increase of
the central SFR caused by the instability.  The SNeII formed in the
central regions dominate the feedback again and enhance the abundance
of $\alpha$-elements. This effect together with the averaging over the
central 2 kpc is also the source of the scatter in the plots for
models D and F.

We note that, because we use an IMF-averaged yield for SNeII, we do
not find a broad range of [O/Fe] values at low metallicities, as would
be expected at early times when bulge stars are enriched by individual
SNeII with different progenitor masses \citep{argast00}. 
The scatter seen for models D and F in
Fig.~\ref{metdistr3} comes from the form of the SFR.

Contrary to the cold model B, the [Mg/Fe] abundance ratio distribution
of the stars in model F (Fig.~\ref{metdistr4}) has only a single peak
around ${\rm [Mg/Fe]} \simeq -0.2$.The different maximum metallicities
reflect the fact that the SF histories in the models are different
(compare Fig.~\ref{massant}).

%
\section{Conclusion}
\label{concl}

We have investigated a sequence of models for star-forming, gas-rich
disks. The most important result from these models is that galaxy
evolution proceeds very differently depending on whether it is the gas
disk or the stellar disk which first becomes unstable, as measured by
the respective Q-parameters. In our two-phase, star-forming
interstellar medium description this depends on how efficiently the
cold cloud medium can dissipate the kinetic energy it gains from
dynamical and feedback heating.
An additional important variable is the infall history \citep{noguchi99},
which influences the gas density and hence the local self-regulation equilibrium
\citep[][SG03]{samland03} and the disk stability.

When the cold gas cools efficiently and drives the instability, the
galactic disk fragments and forms a number of massive clumps. The
stellar disk fragments with the gas because of the strong gravity of
the clumps, which begin to form stars at a high rate because of their
large density. The clumps then spiral to the center of the galaxy in a
few dynamical times and merge there to form a central bulge component
in a strong starburst. This scenario is similar to that discussed by
\citet{noguchi99}.  We show that, in this mode of disk evolution, the
bulge forms rapidly and early; the unstable region of the disk is
completely disrupted, and the disk must be rebuilt by subsequent
infall.  Because of the starburst origin, many of the bulge stars
formed in this way have large [$\alpha$/Fe] abundance ratios.

On the other hand, if the kinetic energy of the cold clouds is
dissipated at a lower rate, stars form from the gas in a more
quiescent mode, while keeping the kinetic temperature high enough and
the gas density low enough to prevent the gas from becoming
dynamically unstable.  In this case, an instability only sets in at
later times, when the surface density of the stellar disk has grown
sufficiently high.  The system then forms a stellar bar, which
channels gas into the center, evolves, and forms a bulge whose stars
are the result of a more extended star formation history, i.e., have
lower [$\alpha$/Fe].  This scenario resembles the evolution described
by \citet{combes81}, \citet{pfenniger90} and \citet{raha91}.  The
comparably long formation time for bars and bulges in this mode and
the weakening of the bar through inward gas flow in the still gas-rich
disks might explain the absence of bars at high redshift
\citep{abraham99}.

An example of a nearby galaxy in which the fragmentation process may
be taking place, is the gas-rich, blue starburst galaxy NGC 7673
\citep{homeier99}. At higher redshifts, a number of
morphologically unusual galaxies may be in similar evolutionary
states, such as the so-called chain galaxies \citep{cowie95} and other
knotty galaxies seen in the HDF \citep{vandenbergh96}; see
\citet{immeli03} for further discussion. On the other hand, the
abundance ratios in the centers of nearby bulges show only a moderate
$\alpha$ overabundance ([$\alpha$/Fe]$ \simeq 0.2$; \citet{proctor02}), 
arguing that these stars were formed from gas enriched by more
extended star formation histories.  The few [Mg/Fe] abundance ratio
measurements available for the Galactic bulge \citep{mcwilliam94}, on
the other hand, point towards more enhanced [$\alpha$/Fe]. Consistent
with this, \citet{puzia02} find a similar
enhancement in three Galactic bulge fields as in bulge globular
clusters from integrated light measurements.

Clearly, mixed star formation histories would also be possible, in
which a part of the bulge formed in an early starburst, preceding a
slower build-up of bulge stars from secular evolution of the disk.  In
the multi-phase galaxy evolution model of SG03, which follows the
formation of a large disk galaxy in a growing $\Lambda$CDM dark matter
halo, the final bulge is indeed a superposition of these two
components with, in that model, the larger part of the mass in the
secular component. Our models suggest that the time-scale for secular
disk evolution is comparable to or longer than that for Fe enrichment
from SNeIa, while the central starburst occurs before SNeIa become
important. Thus further spatially resolved measurements of
[$\alpha$/Fe] in bulges, and measurements of [$\alpha$/Fe] vs Fe and
radius for Galactic bulge stars will be key for disentangling bulge
formation history.

It is well-known that the properties and evolution of a galactic disk
depend on global parameters like mass, angular momentum, and
infall rate.  We have shown here that the evolution may also depend
strongly on the physics of the baryonic component, in particular, the
uncertain energy dissipation rate of the cold cloud medium from which
most stars are formed. This also implies that one cannot simply derive
the accretion rate onto a galaxy by determining its star formation
history.  Although all the models in our sequence have an identical
gas infall, we observe very different SF histories from one model to
the other.  In all models, however, the instability causes gas
transport to the center, and a subsequent increase of the central SFR
leading to the build-up of a central bulge component.

\vspace{0.5cm} 
We thank the referee, M. Noguchi, for his prompt report, the
Schweizerischer Nationalfond for financial support of this work under
grant 20-64856.01, and the Centro Svizzero di Calcolo Scientifico
(CSCS) for the opportunity of using their computing facilities.



\begin{thebibliography}{}

\bibitem[Abadi et al.(2003)]{abadi03}
        Abadi, M.~G., Navarro, J.~F., Steinmetz, M., Eke, V.~R. 2003, \apj, 591, 499

\bibitem[Abraham et al.(1996)]{abraham96}
        Abraham, R.~G., Tanvir, N.~R., Santiago, B.~X., Ellis, R.~S, Glazebrook, K.,  
        van den Bergh, S. 1996, \mnras, 279, L47

\bibitem[Abraham et al.(1999)]{abraham99}
        Abraham, R.~G., Merrifield, M.~R., Ellis, R.~S., Tanvir, N.~R., 
        Brinchmann, J. 1999, \mnras, 308, 569

\bibitem[Abraham(1999)]{abraham99b}
        Abraham, R.~G. 1999, \apss, 269, 323

\bibitem[Abraham \& Merrifield(2000)]{abraham00}
        Abraham, R. G., Merrifield, M. R. 2000, \aj, 120, 2835


\bibitem[Aguerri(1999)]{aguerri99}
        Aguerri, J. A. L. 1999, A\&A, 351, 43

\bibitem[Aguerri et al.(2001)]{aguerri01}
        Aguerri, J. A. L., Balcells, M., Peletier, R. F.  2001, A\&A, 367, 428


\bibitem[Argast et al.(2000)]{argast00}
        Argast, D., Samland, M., Gerhard, O., Thielemann, F.-K. 2000, A\&A, 356, 873

\bibitem[Balsara et al.(2001)]{balsara01}
        Balsara, D., Ward-Thompson, D., Crutcher, R. M. 2001, \mnras, 327, 715


\bibitem[Brinchmann et al.(1998)]{brinchmann98}
        Brinchmann, J., Abraham, R., Schade, D., Tresse, L.,
        Ellis, R. S., Lilly, S., Le Fevre, O., Glazebrook, K.,
        Hammer, F., Colless, M., Crampton, D., Broadhurst, T. 1998, \apj, 499, 112

\bibitem[Bullock et al.(2001)]{bullock01}
        Bullock, J. S., Dekel, A., Kolatt, T. S., Kravtsov, A. V., Klypin, A. A., 
        Porciani, C., Primack, J. R. 2001 \apj, 555, 240

\bibitem[Cha \& Whitworth(2003)]{cha03}
        Cha, S.-H., Whitworth, A. P. 2003, \mnras, 340, 91

\bibitem[Chen \& Jing(2002)]{chen02}
        Chen, D. N., Jing, Y. P. 2002, \mnras, 336, 55

\bibitem[Chiba \& Beers(2000)]{chiba00}
        Chiba, M., Beers, T. 2000, \aj, 119, 2843

\bibitem[Cole et al.(2000)]{cole00}
        Cole, S., Lacey, C. G., Baugh, C. M., Frenk, C. S. 2000, \mnras, 319, 168

\bibitem[Combes \& Sanders(1981)]{combes81}
        Combes, F., Sanders, R. H. 1981, A\&A 96, 164

\bibitem[Contardo et al.(1998)]{contardo98}
        Contardo, G., Steinmetz, M., Fritze-v. Alvensleben, U. 1998, \apj, 507, 497

\bibitem[Courteau et al.(1996)]{courteau96}
        Courteau, S., de Jong, R. S., Broeils, A. H. 1996, \apjl, 457, 73

\bibitem[Cowie et al.(1995)]{cowie95}
        Cowie, L. L., Hu, E. M., Songaila, A. 1995, \aj, 110, 1576 (CHS95)

\bibitem[Cowie et al.(1980)]{cowie80}
        Cowie, L. L. 1980, \apj, 236, 868

\bibitem[Das et al.(2003)]{das03}
        Das, M., Teuben, P. J., Vogel, S. N., Regan, M. W., Sheth, K., Harris, A. I.,
        Jefferys W. H. 2003, \apj, 582, 190

\bibitem[Dickinson(2000)]{dickinson00}
        Dickinson, M. 2000, in ``Building Galaxies: From the Primordial Universe 
        to the Present'', XIXth Moriond Astrophysics Meeting, ed. F. Hammer et al. 
        (Paris: Ed. Frontieres), 257

\bibitem[Driver et al.(1998)]{driver98}
        Driver, S. P., Fernandez-Soto, A., Couch, W. J., Odewahn, S. C.,
        Windhorst, R. A., Phillips, S., Lanzetta, K., Yahil, A. 1998, \apj, 496, 93

\bibitem[Durrell et al.(2001)]{durrell01}
        Durell, P. R., Harris W. E., Pritchet, C. J. 2001, \apj, 121, 2557

\bibitem[Edvardsson et al.(1993)]{edvardsson93}
        Edvardsson, B., Andersen, J., Gustafsson, B., Lambert, D. L., Nissen, P. E.,
        Tomkin, J. 1993, A\&A, 102, 603

\bibitem[Eggen et al.(1962)]{eggen62}
        Eggen, O. J., Lynden-Bell, D., Sandage, A. R. 1962, \apj, 136, 748

\bibitem[Elmegreen(1995)]{elmegreen95}
        Elmegreen, B.G. 1995, \mnras, 275, 944

\bibitem[Elmegreen(1989)]{elmegreen89}
        Elmegreen, B.G., 1989, \apj, 338, 178

\bibitem[Friedli \& Benz(1993)]{friedli93}
        Friedli, D., Benz, W. 1993, A\&A, 268, 65


\bibitem[Gingold \& Monaghan(1983)]{gingold83}
        Gingold, R. A., Monaghan, J. J. 1983, \mnras, 204, 715

\bibitem[Guiderdoni et al.(1998)]{guiderdoni98}
        Guiderdoni, B., Hivon, E., Bouchet, F. R., Maffei, B. 1998, \mnras, 295, 877

\bibitem[Homeier \& Gallagher(1999)]{homeier99}
        Homeier N. L., Gallagher J. S., 1999, \apj, 522, 199

\bibitem[Immeli et al.(2003)]{immeli03}
        Immeli, A., Samland, M., Gerhard, O.E., Westera, P. 2003, \apjl, submitted

\bibitem[Jenkins et al.(2001)]{jenkins01}
        Jenkins, A., Frenk, C. S., White, S.~D.~M., Colberg, J. M., Cole, S., 
        Evrard, A. E., Couchman, H.~M.~P., Yoshida, N. 2001, \mnras, 321, 372

\bibitem[Jog \& Solomon(1984)]{jog84}
        Jog, C. J., Solomon, P. M., 1984, \apj, 276, 114

\bibitem[Kauffmann et al.(1993)]{kauffmann93}
        Kauffmann, G., White, S.~D.~M., Guiderdoni, B. 1993, \mnras, 264, 201

\bibitem[Kajisawa \& Yamada(2001)]{kajisawa01}
        Kajisawa, M., Yamada, T. 2001, \pasj, 53, 833

\bibitem[Kennicutt(1998)]{kennicutt98}
        Kennicutt, R. C., 1998, \apj, 498, 541

\bibitem[Kim et al.(2001)]{kim01}
        Kim, J., Balsara, D., Mac Low, M.-M. 2001, JKAS, 34, 333

\bibitem[Kormendy(1982)]{kormendy82}
        Kormendy, J. 1982, in {\sl Morphology and Dynamics of Galaxies}, ed.
        L. Martinet \& M. Mayor (Sauverny: Geneva Obs.), 113

\bibitem[Klypin et al.(2001)]{klypin01}
        Klypin, A., Kravtsov, A. V., Bullock, J. S., Primak, J. R. 2001, \apj, 554, 903

\bibitem[Larson(1969)]{larson69}
        Larson, R.B., 1969, MNRAS, 145, 405

\bibitem[Lilly et al.(1998)]{lilly98}
        Lilly, S., Schade, D., Ellis, R., Le Fevre, O.,
        Brinchmann, J., Tresse, L., Abraham, R., Hammer,
        F., Crampton, D., Colless, M., Glazebrook, K.,
        Mallen-Ornelas, G., Broadhurst, T. 1998, \apjl, 500, 75L

\bibitem[McKee \& Ostriker(1977)]{mckee77}
        McKee, C. F., Ostriker, J. P., 1977, ApJ, 218, 148

\bibitem[McWilliam \& Rich(1994)]{mcwilliam94}
        McWilliam, A., Rich, R. M. 1994, \apjs, 91, 749

\bibitem[Moore et al.(1998)]{moore98}
        Moore, B., Governato, F., Quinn, T., Stadel, J., Lake, G. 1998, \apjl, 499, 5

\bibitem[Navarro et al.(1996)]{navarro96}
        Navarro, J. F., Frenk, C. S., White, S. D. M. 1996, \apj, 462, 563

\bibitem[Navarro et al.(1997)]{navarro97}
        Navarro, J. F., Frenk, C. S., White, S. D. M. 1997, \apj, 490, 493

\bibitem[Navarro \& Steinmetz(1997)]{navarro97b}
        Navarro, J. F., Steinmetz, M., 1997, \apj, 478, 13

\bibitem[Noguchi(1999)]{noguchi99}
        Noguchi, M. 1999, \apj, 514, 77

\bibitem[Norman et al.(1996)]{norman96}
        Norman, C. A., Sellwood, J. A., Hasan, H. 1996, \apj, 462, 114

\bibitem[O'Neil et al.(2000)]{oneil00}
        O'Neil, K., Bothun, G. D., Impey, C. D., 2000 \apjs, 128, 99

\bibitem[Peletier \& Balcells(1996)]{peletier96}
        Peletier, R. F., Balcells, M. 1996, AJ, 111, 2238

\bibitem[Pfenniger \& Norman(1990)]{pfenniger90}
        Pfenniger, D., Norman, C. 1990, \apj, 363, 391

\bibitem[Polyachenko(1997)]{polyachenko97}
        Polyachenko, V. L, Polyachenko, E. V., Strel'Nikov, A. V. 1997, AstL, 23, 483

\bibitem[Puzia et al.(2002)]{puzia02} 
        Puzia, T. H., Saglia, R. P., Kissler-Patig, M., Maraston, C., Greggio, L., 
        Renzini, A., Ortolani, S. 2002, A\&A, 395, 45

\bibitem[Raha et al.(1991)]{raha91}
        Raha, N., Sellwood, J. A., James, R. A., Kahn, F. D. 1991, Nature 352, 411

\bibitem[Ramirez et al.(2000)]{ramirez00}
        Ramirez, S. V., Stephens, A. W., Frogel, J. A., DePoy, D. L. 2000, \aj, 120, 833

\bibitem[Romeo(1992)]{romeo92}
        Romeo, B.R. 1992, \mnras, 256, 307

\bibitem[Safranov(1960)]{safranov60}
        Safranov, V.S. 1960, Ann. d'Ap., 23, 979

\bibitem[Samland(1998)]{samland98}
        Samland, M. 1998, \apj, 496, 155

\bibitem[Samland \& Gerhard(2003)]{samland03}
        Samland, M., Gerhard, O.  2003, A\&A, 399, 961 (SG03)

\bibitem[Proctor \& Sansom(2002)]{proctor02}
        Proctor, R. N., Sansom, A. E. 2002, \mnras, 333, 517

\bibitem[Scannapieco \& Tissera(2003)]{scannapieco03}
        Scannapieco, C., Tissera, P. B. 2003, \mnras, 338, 880

\bibitem[Schmidt(1959)]{schmidt59}
        Schmidt, M. 1959, \apj, 129, 243

\bibitem[Seigar et al.(2002)]{seigar02}
        Seigar, M., Carollo, C. M., Stiavelli, M., de Zeeuw, P. T., Dejonghe, H.
        2002, \aj, 123, 184

\bibitem[Sommer-Larsen et al.(2002)]{sommer02}
        Sommer-Larsen, J., G{\"o}tz, M., \& Portinari, L. 2002, astro-ph/0204366

\bibitem[Sommer-Larsen et al.(1999)]{sommer99}
        Sommer-Larsen, J., Gelato, S., Vedel, H. 1999, \apj, 519, 501

\bibitem[Steinmetz \& M\"uller(1995)]{steinmetz95}
        Steinmetz, M., M\"uller, E. 1995, \mnras, 276, 549

\bibitem[Tiede \& Terndrup(1999)]{tiede99}
        Tiede, G. P., Terndrup, D. M. 1999, \aj 118, 895

\bibitem[Toomre(1964)]{toomre64}
        Toomre, A. 1964, \apj, 139, 1217

\bibitem[van den Bergh et al.(1996)]{vandenbergh96}
        van den Bergh, S., Abraham, R.G., Ellis, R.S., Tanvir, N.R.,
        Santiago, B.X., Glazebrook, K.G., 1996, \aj, 112, 359

\bibitem[van den Bergh et al.(2002)]{vdb02}
        van den Bergh, S., Abraham, R. G., Whyte, L. F., Merrifield, M. R.,
        Eskridge, P. B., Frogel J. A., Pogge, R. 2002, \aj, 123, 2913

\bibitem[van den Bergh et al.(2000)]{vandenbergh00}
        van den Bergh, S., Cohen, J.G., Hogg, D.W., Blandford, R. 2000, \aj, 120, 2190

\bibitem[van den Bosch(2002)]{vdbosch02}
        van den Bosch, F. C. 2002, \mnras, 332, 456

\bibitem[Wang \& Silk(1994)]{wang94}
        Wang, B., Silk, J. 1994, \apj, 427, 759

\bibitem[Wechsler et al.(2002)]{wechsler02}
        Wechsler, R. H., Bullock, J. S., Primack, J. R., Kravtsov, A. V., Dekel, A.
        2002, ApJ 568, 52  

\bibitem[Westera et al.(2002)]{westera02}
        Westera, P., Samland, M., Gerhard, O., Buser, R. 2002, A\&A, 389, 761

\bibitem[Williams \& Nelson(2001)]{williams01}
        Williams, P. R., Nelson, A. H. 2001, A\&A, 374, 839

\bibitem[Zoccali et al.(2003)]{zoccali03}
        Zoccali, M., Renzini, A., Ortolani, S., Greggio, L., Saviane, I., Cassisi, S.,
        Rejkuba, M., Barbuy, B., Rich, R. M., Bica, E. 2003, A\&A, 399, 931
\end{thebibliography}
\end{document}